  \documentclass{emulateapj}

  \usepackage{graphicx}
\def\ltwig{\mathrel{\spose{\lower 3pt\hbox{$\mathchar"218$}}
       \raise 2.0pt\hbox{$\mathchar"13C$}}}
\def\gtwig{\mathrel{\spose{\lower 3pt\hbox{$\mathchar"218$}}
       \raise 2.0pt\hbox{$\mathchar"13E$}}}
\def\spose#1{\hbox to 0pt{#1\hss}}

\newcommand{\beq}{\begin{equation}}
\newcommand{\eeq}{\end{equation}}
\newcommand{\beqa}{\begin{eqnarray}}
\newcommand{\eeqa}{\end{eqnarray}}

\def\=={{\equiv}}
\def\Rstar{R_{\ast}}
\def\Mstar{M_{\ast}}
\def\Lstar{L_{\ast}}

\begin{document}

\title{A Nozzle Analysis of Slow-Acceleration Solutions in \\
One-Dimensional Models of Rotating Hot-Star Winds}
\shorttitle{Nozzle Analysis of Rotating Hot-Star Winds}

\email{tmadura@udel.edu\\ owocki@bartol.udel.edu\\
afeld@astro.physik.uni-potsdam.de}

\author{Thomas I. Madura and Stanley P. Owocki}
\affil{Bartol Research Institute, Department of Physics \& Astronomy,
    University of Delaware, Newark, DE 19716}
\author{Achim Feldmeier}
\affil{Astrophysik, Institut f\"{u}r Physik, Universit\"{a}t
Potsdam, Am Neuen Palais 10, 14469 Potsdam, Germany}

\begin{abstract}
One-dimensional (1D) stellar wind models for hot stars rotating at
$\gtrsim$ 75\% of the critical rate show a sudden shift to a
slow-acceleration mode, implying a slower, denser equatorial outflow
that might be associated with the dense equatorial regions inferred
for B[e] supergiants. Here we analyze the steady 1D flow equations
for a rotating stellar wind based on a ``nozzle'' analogy for terms
that constrain the local mass flux. For low rotation, we find the
nozzle minimum occurs near the stellar surface, allowing a
transition to a standard, CAK-type steep-acceleration solution; but
for rotations $\gtrsim$ 75\% of the critical rate, this inner nozzle
minimum exceeds the global minimum, implying near-surface
supercritical solutions would have an overloaded mass loss rate. In
steady, analytic models in which the acceleration is assumed to be
monotonically positive, this leads the solution to switch to a slow
acceleration mode.
However, time-dependent simulations using a numerical hydrodynamics
code show that, for rotation rates 75 - 85\% of critical, the flow
can develop abrupt ``kink" transitions from a steep acceleration
to a {\em decelerating} solution. For rotations above 85\% of
critical, the hydrodynamic simulations confirm the slow
acceleration, with the lower flow speed implying densities 5 - 30
times higher than the polar (or a nonrotating) wind. Still, when
gravity darkening and 2D flow effects are accounted for, it seems
unlikely that rotationally modified equatorial wind outflows could
account for the very large densities inferred for the equatorial
regions around B[e] supergiants.
\end{abstract}


\section{Introduction} \label{sec:intro}

Hot, luminous, massive stars of spectral types O and B are generally
quite rapid rotators, with inferred surface rotation speeds
typically in the range of several 100~km/s, or a substantial
fraction of the critical rotation speed (ca. 400-600~km/s) at which
material at the rotating equatorial surface would be in Keplerian
orbit. Such luminous stars are also characterized by strong stellar
wind outflows, driven by the line-scattering of the star's continuum
radiation flux \citep[ hereafter CAK]{CAK}. A longstanding question
is how such wind outflows are affected by the star's  rotation, and
in particular, whether this might play a role in the enhanced
equatorial density outflows and/or disks inferred in certain classes
of particularly rapid rotators, e.g., Be and B[e] stars.

For classical Be stars, there is now substantial observational
evidence \citep[see, e.g., papers in][]{IG05} that the disks are
Keplerian in nature, with very limited radial outflow; they are thus
probably \emph{not} a direct result of feeding by a steady stellar
wind \citep{Owocki05}. However, for the disk and/or enhanced
equatorial outflows inferred in supergiant B[e] stars, wind
mechanisms seem still a viable option. For example, \citet{LP91} and
\citet{Pelupetal00} have noted that the ``bi-stability'' enhancement
in opacity which occurs for some value of the local surface
effective temperature (i.e. for B-type stars at $\sim 22,000$ K, see
\citet{PualPuls, Vink}) associated with the reduced radiation
temperature near the equator can lead to a factor several
enhancement in the radial mass flux. By itself, this seems
inadequate to explain equatorial densities estimated to be hundreds
or even thousands of times the densities of the polar winds in these
stars \citep{Zickgrafetal85, KrausMiro06}.
%
%
However, a recent series of papers by M. Cur{\'e} and colleagues
\citep{Cure04, CureRial04, Cureetal05} proposes that, for very
high, near-critical rotation, a switch of the wind outflow to a
slower, shallow-acceleration solution can lead to a further
enhancement in density that, together with the bi-stability
effect, might reach the equatorial densities inferred in B[e]
supergiants.
%

The present paper aims to understand better the physical origin of
these shallow wind acceleration solutions for high rotation rates, and
to examine critically their likely relevance for explaining dense equatorial
disks or outflows.


Modelling rotating, hot-star winds began with the studies by \citet[
hereafter FA]{FA} and by \citet[ hereafter PPK]{PPK}, who extended
the CAK formalism by adding the effect of an outward centrifugal
acceleration to one-dimensional (1D) models of the wind outflow in
the equatorial plane.
Both FA and PPK independently derived a modified CAK model (``mCAK")
that relaxes the CAK ``point-star" approximation and properly
accounts for the finite cone-angle subtended by the star.
They then each found that the reduction in the effective gravity by
the outward centrifugal force tends to increase the mass-loss rate
and decrease the wind speed. However, for the models computed, up to
about 75\% of the critical rotation rate, both changes are limited
to only a factor of a few and are thus insufficient to produce the
large equatorial densities and low velocities inferred in B[e]
supergiants. Moreover, for still faster rotation, above about 75\%
critical, FA found that the equations for outward acceleration could
no longer be integrated beyond some finite radius, and thus they
were unable to derive any complete flow solutions for such rapid
rotation speeds.
Subsequent 1D models have investigated the role of magnetic forces
\citep{FriendMacGreg84, PoeFriend86} and sound-waves \citep{Koninx92},
but as summarized by \citet{BC93} neither mechanism seems favorable
to producing slow, dense equatorial outflows.

More fundamentally, the physical relevance of any such 1D models
may be limited since accounting for latitudinal flows toward or
away from the equator requires at least a two-dimensional (2D)
treatment. A vivid example comes from the 2D ``Wind Compressed
Disk" (WCD) model of \citet{BC93}, which argues that conservation
of angular momentum should tend to channel material from higher
latitudes toward an equatorially compressed disk flow. If one
assumes a purely radial driving force, 2D hydrodynamical
simulations
\citep{Owockietal94} confirm the basic WCD effect, but show that,
depending on whether material reaches the equator above or below
some  ``stagnation point", it either drifts outward or falls back
toward the star. Such simultaneous infall-plus-outflow behavior is
not possible in a steady 1D model, but is a perfectly natural
outcome in a 2D simulation. Furthermore, \citet{CO95} showed that,
when computed from proper angle integration of intensity from the
rotationally distorted stellar surface, the line-force also has
nonzero, \emph{nonradial} components in both  azimuth and
latitude. \citet[][ see also \citet{PP00}]{Owockietal96} showed
that such nonradial line forces can \emph{inhibit} the formation
of a WCD. Finally, equatorial gravity darkening can actually
\emph{reduce} the wind mass flux from the equator, and so lead to
an equatorial wind density that is \emph{lower}, not higher, than
near the poles.

Despite this likely importance of such 2D effects,
several recent analyses \citep{Cure04, CureRial04, Cureetal05}
have reexamined the 1D equatorial flow models of FA,
with a particular focus on the failure to obtain monotonically
accelerating wind solutions for rotation above about 75\% of the critical speed.
In particular, these papers argue that for such very high, near-critical
rotation, the wind solution can switch to an alternative mode,
characterized by a much {\it slower} outward acceleration.
Together with a moderately enhanced mass flux, the resulting lower-speed
outflow then implies a substantial enhancement in density,
relative to the standard CAK, steep acceleration applicable at higher
latitudes.
When combined with parameterizations intended to mimic a
``bi-stability'' enhancement in the line-driving \citep{LP91,
Pelupetal00},
\citet{Cureetal05} predict equator-to-pole density contrasts of the order of
$10^{2}-10^{4}$.

Assessing the physical relevance of such claims for understanding B[e]
stars will eventually require proper account of the multi-dimensional
effects noted above.
Nonetheless, to provide a solid basis for such multi-dimensional
models, it is important to have a clearer, dynamical understanding
of these novel 1D slow outflow solutions.
%
%
Combining analytic studies with numerical hydrodynamic simulations,
this paper examines the reality and physical origin of the shallow
wind acceleration solutions for high rotation rates, with emphasis
on their possible relevance to disk outflows from B[e] supergiants.

We begin (\S\,\ref{sec:Massloss}) with a basic review of the
general time-dependent wind equations, together with their
CAK-type, steady-state solutions in a non-rotating wind. To
provide a physical basis for extending these steady models to
include rotation, we first (\S\,\ref{sec:SSsolns}) apply a simple
``nozzle" analysis \citep[see][]{Holzer77, Abbott1980}, originally
developed to study winds driven from luminous accretion disks
\citep{Pereyraetal04}. With a few judicious, yet quite reasonable
approximations (e.g., neglecting gas pressure terms by taking the
zero-sound-speed limit; using a beta velocity law to evaluate the
finite-disk correction as an explicit function of radius), it is
possible to obtain simple integrations of the equation of motion
and to study the scalings of the mass loss rate with rotation, as
well as the switch from steep to shallow acceleration solutions
beyond a threshold rotation rate. To test the validity of these
simple analytic solutions, we next apply a numerical hydrodynamics
code (\S\,\ref{subsec:HydroSimSpecs}) to evolve 1D rotating  wind
models to asymptotic steady states (\S\,\ref{subsec:AsympState}).
Results confirm a transition to slower acceleration at very high
rotation (above about 85\% of critical), but also show a new class
of non-monotonic ``kink'' solutions that apply for moderately fast
rotation (ca. 75-85\% of critical). We then examine
(\S\,\ref{subsec:TimRelax}-\ref{subsec:varyIC}) the time evolution
of solutions in various rotation domains, with emphasis on the
kink solutions, and on a peculiar transition case (85\% of
critical rotation), characterized by an initial wind overloading
followed by a flow stagnation and eventual reaccretion of material
onto the star. We conclude with a summary and outlook for future
work (\S\,\ref{sec:summary}).



\section{General Formalism for Line-Driven Mass-Loss}
\label{sec:Massloss}

\subsection{1D Time-Dependent Equations of Motion} \label{subsec:GenEoM}

In this paper, we examine 1D radiatively driven outflow in the equatorial plane
of a rotating  star.
For the general time-dependent simulations discussed in \S 3, the
relevant  equations for conservation of mass and radial component
of momentum have the form
\beq
\frac{\partial \rho}{\partial t} + \frac{1}{r^{2}}
\frac{\partial(r^{2}\rho v)}{\partial r} = 0 \,
\label{hydro1}
\eeq
\beq \frac{\partial v}{\partial t} + v\frac{\partial v}{\partial r}
= \frac{v_{\phi}^{2}}{r} - \frac{1}{\rho}\frac{\partial P}{\partial
r} - {GM_{\ast} (1-\Gamma_{e}) \over r^{2}} + g_{\rm lines}, \,
\label{hydro2} \eeq where $r$ and $t$ are the radius and time, and
$\rho$ and $v$ are  the mass density and radial component of the
velocity. The body forces here include the outward radiative
acceleration from line-scattering, $g_{\rm lines}$, and an effective
inward gravitational acceleration $GM_{\ast}(1-\Gamma_{e})/r^{2}$,
reduced by the outward continuum radiative force from scattering by
free electrons, as accounted for by the Eddington parameter
$\Gamma_{e} = \kappa_{e} L_{\ast}/4\pi GM_{\ast} c$.
For the centrifugal term, $v_{\phi}^{2}/r$, we avoid explicit
treatment of an azimuthal momentum equation by assuming simple
angular momentum conservation (which is a good approximation in
the supersonic flow domain considered here), yielding then for the
azimuthal speed \beq v_{\phi} = v_{\rm rot} \, { R_{\ast} \over r
} \, , \label{vphidef} \eeq where $v_{\rm rot}$ is the rotation
speed at the star's equatorial surface radius $R_{\ast}$. For
simplicity, we also avoid a detailed treatment of the wind energy
balance by assuming an isothermal outflow, for which the pressure
is written as $P = \rho a^{2}$, where $a$ is the (constant,
isothermal) sound speed.

\subsection{Steady-State Equations with Rotation} \label{subsec:SSEoM}

For the simplified case of a steady state, the time-dependent terms
vanish ($\partial/\partial t = 0$), yielding for the steady
acceleration \beq v\frac{dv}{dr} =
-\frac{GM_{\ast}(1-\Gamma_{e})}{r^{2}} + {v_{\rm rot}^{2}
R_{\ast}^{2} \over r^{3}} + g_{\rm lines} -
\frac{a^{2}}{\rho}\frac{d\rho}{dr} \, . \label{genEoM} \eeq The
steady form for mass conservation implies a constancy for the
overall mass loss rate, $\dot{M} \equiv 4 \pi \rho v r^{2}$. Using
this to eliminate the density $\rho$, the equation of motion
(\ref{genEoM}) takes the form \beq \left [1 - \frac{a^{2}}{v^{2}}
\right ]v\frac{dv}{dr} = -\frac{GM_{\ast}(1-\Gamma_{e})}{r^{2}} +
{v_{\rm rot}^{2} R_{\ast}^{2} \over r^{3}} + g_{\rm lines} +
\frac{2a^{2}}{r}
\, . \label{EoM} \eeq The square-bracket factor on the
left-hand-side allows for a smooth mapping of the wind base onto a
hydrostatic atmosphere below the sonic point, where $v < a$.
However, in radiatively driven winds the pressure terms on the
right-hand-side are generally negligible since, compared to the
gravitational acceleration term that must be overcome to drive a
wind, these are of order $w_{s} \equiv (a/v_{\rm esc})^{2} \approx
0.001$, where $v_{\rm esc} \equiv
\sqrt{2GM_{\ast}(1-\Gamma_{e})/R_{\ast}}$ is the effective escape
speed from the stellar surface radius $R_{\ast}$.

Since the key to a stellar wind is to overcome gravity, it is
convenient to define a dimensionless equation of motion that
scales all accelerations by the effective gravity, \beq (1 -
w_{s}/w)w' = -1 + \omega^{2} (1-x) + \Gamma_{\rm lines} +
\frac{4w_{s}}{(1-x)} \, , \label{ScaledEoM} \eeq where
$\Gamma_{\rm lines} \equiv g_{\rm
lines}r^{2}/GM_{\ast}(1-\Gamma_{e})$, and the gravitationally
scaled inertia is $w' \equiv dw/dx  =
r^{2}v(dv/dr)/GM_{\ast}(1-\Gamma_{e})$. The independent variable
here is the inverse radius coordinate $x \equiv 1 - \Rstar/r$,
while the dependent variable is the ratio of the radial kinetic
energy to effective surface escape energy, $w \equiv v^2/v_{\rm
esc}^{2}$. Gas pressure effects are accounted for by terms
containing
$w_{s} \equiv a^{2}/v_{\rm esc}^{2}$, while centrifugal effects
from rotation are characterized in terms of the ratio of the
equatorial rotation speed to critical speed, $\omega \equiv v_{\rm
rot}/v_{\rm crit} = \sqrt{2} \, v_{\rm rot}/v_{\rm esc}$, under
the assumption that the wind material conserves its surface value
of specific angular momentum, $r v_{\phi}(r) = v_{\rm rot}
\Rstar$.

Within the CAK formalism for driving by scattering of a
point-source of radiation by an ensemble of lines,
the deshadowing of optically thick lines by the Doppler shift
associated with the wind acceleration gives the scaled radiative
acceleration $\Gamma_{\rm lines}$ a power-law dependence on the
flow acceleration $w'$, \beq \Gamma_{\rm lines} =  C \,
w'^{\alpha} \, , \label{GammaCAK} \eeq where $\alpha$ is the CAK
power index. Here we have eliminated an inverse dependence on
density $\rho$ in favor of the mass loss rate $\dot{M} = 4\pi
r^{2} \rho v$,
with the line force constant thus defined by \beq C \equiv
\frac{1}{1 - \alpha} \left [\frac{L_{\ast}}{\dot{M}c^{2}} \right
]^{\alpha} \left[{ \bar{Q} \Gamma_{e} \over 1- \Gamma_{e}}
\right]^{1-\alpha} \, , \label{C} \eeq with $L_{\ast}$ the stellar
luminosity, and $\Gamma_{e}$ the Eddington parameter for the
gravitationally scaled radiative acceleration from electron
scattering opacity, $\kappa_{e}$ (cm$^{2}$/g) \citep{LC, Oleron}. We
have also used the \citet{Gayley95} $\bar{Q}$ notation for the
overall normalization of the line opacity. Note that, for fixed sets
of stellar parameters ($L_{\ast}, M_{\ast}, \Gamma_{e}$) and
line-opacity ($\alpha, \bar{Q}$), the constant $C$ scales with the
mass loss rate as $C \propto 1/\dot{M}^{\alpha}$.

As already noted, the smallness of the dimensionless sound-speed
parameter $w_{s}$ implies that gas pressure plays little role in
the dynamics of any line driven stellar wind. Hence, to a good
approximation, we can obtain accurate solutions by analyzing the
much simpler limit of vanishing sound speed $a \propto
\sqrt{w_{s}} \rightarrow 0$, for which the line-driven-wind
equation of motion reduces to \beq w' = - 1 + \omega^{2} (1-x) + C
w'^{\alpha} \, . \label{dimlessEoM} \eeq

\subsection{Classical CAK Solution for a Point-Star} \label{subsec:CAKsoln}

Let us first review the standard CAK solution without rotation,
setting $\omega=0$. Note then that since the parameters $\Gamma_{e}$
and $C$ are spatially constant, the solution is independent of
radius. For high $\dot{M}$ and small $C$ there are no solutions,
while for small $\dot{M}$ and high $C$ there are two solutions. The
CAK \emph{critical} solution (denoted by the subscript $c$)
corresponds to a \emph{maximal} mass loss rate, which requires a
\emph{tangential} intersection between the line-force $C
w'^{\alpha}$ and the combined inertia plus gravity $1+w'$, for which
\beq \alpha C_{c} w_{c}'^{\alpha-1} = 1 \, , \eeq and thus, together
with the equation of motion (\ref{dimlessEoM}), we have \beq w_{c}'
= \frac{\alpha}{1 - \alpha} \, , \eeq with \beq C_{c} =
\frac{1}{\alpha^{\alpha} \left ({1 - \alpha} \right )^{1-\alpha} }
\, . \label{Cc} \eeq Using eq. (\ref{C}), this then yields the
standard CAK scaling for the mass loss rate, \beq \dot{M}_{\rm CAK}
= \frac{L_{\ast}}{c^{2}} \frac{\alpha}{1- \alpha} \left
[\frac{\bar{Q}\Gamma_{e}}{1 - \Gamma_{e}}
\right]^{(1-\alpha)/\alpha} \, . \label{MdotCAK} \eeq Moreover,
since the scaled equation of motion (\ref{dimlessEoM}) has no
explicit spatial dependence, the scaled critical acceleration
$w_{c}'$ applies throughout the wind. This can therefore be
trivially integrated to yield \beq w(x) = w(1)x \, , \eeq where
$w(1) = \alpha / (1-\alpha)$ is the terminal value of the scaled
flow energy. In terms of dimensional quantities, this represents a
specific case of the general ``beta"-velocity-law, \beq v(r) =
v_{\infty} \left (1 - \frac{R_{\ast}}{r} \right )^{\beta} \, ,
\label{betalaw} \eeq where in this case $\beta = 1/2$, and the wind
terminal speed scales with the effective escape speed from the
stellar surface, $v_{\infty} = v_{\rm esc}\sqrt{\alpha/(1-\alpha)}$.

\subsection{Finite-Disk Form for the CAK Line-Force}
\label{subsec:FDCF}

The above analysis is based on the idealization of radially
streaming radiation, as if the star were a point source at the
origin. This was the basis of the original CAK wind solutions,
although they did already identify (but did not implement) the
appropriate ``finite-disk correction factor" (fdcf) to account for
the full angular extent of the star (see CAK eq. [50 ]), \beq
 f(r)
= \frac {(1+\sigma)^{1+\alpha}-(1+ \sigma \mu_\ast^2 )^{1+\alpha}}
{(1+\alpha) \sigma (1+\sigma)^\alpha (1-\mu_\ast^2)} \, ,
\label{fdcf} \eeq with $\mu_{\ast} \equiv
\sqrt{1-\Rstar^{2}/r^{2}}$ the cosine of the finite-cone angle of
the stellar disk, and $\sigma \equiv d\ln v/d \ln r -1 $. When
this factor is included to modify the point-star CAK line
acceleration (from eq. [\ref{GammaCAK}]),
its complex dependence on radius, velocity, and velocity gradient
complicates the solution of the full equation of motion. Full
solutions derived independently by FA and PPK yield a somewhat
reduced mass-loss rate $\dot{M}_{\rm fd} \approx \dot{M}_{\rm
CAK}/(1+\alpha)^{1/\alpha}$ and higher terminal speed $v_{\infty}
\approx 3v_{\rm esc}$.

\begin{figure}
\begin{center}
\plotone{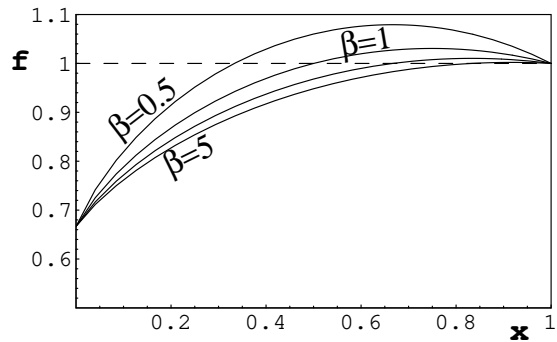} \caption{ Spatial variation of the
finite-disk-correction factor $f$, plotted vs. scaled inverse radius
$x = 1 - R_{\ast}/r$, for  CAK exponent $\alpha= 1/2$ and various
velocity-law exponents, $\beta =$ 0.5, 1, 2, and 5. The horizontal
dashed curves denote the unit correction that applies at the point
of isotropic expansion (where $\sigma = d\ln v/d\ln r - 1 = 0$), and
at large distances where the star approaches the point-source form
assumed in the original CAK model.} \label{fig1}
\end{center}
\end{figure}

But if we approximate the wind velocity law  by the simple
``beta-law'' form of eq. (\ref{betalaw}), then the fdcf can be
evaluated as an \emph{explicit} spatial function. Figure
\ref{fig1} illustrates the resulting variation of $f$ with the
scaled coordinate $x$ for $\alpha=1/2$ and various values of
$\beta$. Note that the overall form is quite similar for all
cases, increasing from a surface value $f_{\ast} \equiv
f(\Rstar)=1/(1+\alpha)$ to past unity at the isotropic expansion
radius (where $dv/dr=v/r$), $r/\Rstar = (1+\beta)$ [corresponding
to $x=\beta/(1+\beta)$], and eventually returning asymptotically
to unity from above at large radii ($x \rightarrow 1$).

In the steady wind analysis in the next section, we thus choose
the canonical value $\beta=1$ to represent the fdcf as an explicit
spatial function (PPK).

\section{Steady-State Solutions for 1D Models of Rotating,
Line-Driven Stellar Winds}
\label{sec:SSsolns}

\subsection{Nozzle Analysis for Steady Wind Acceleration}
\label{subsec:windaccel}

Let us now examine how the combined effects of the fdcf and rotation
alter the classical CAK result. Note that we are ignoring here
gravity darkening and oblateness effects, as well as any
``bi-stability'' in the line-driving parameters between the polar
and equatorial wind. For a rotating star and wind, the fdcf can
become even more complicated, modified by the rotational shear of
the wind outflow, and by the oblateness of the star, and possibly
also by the equatorial gravity darkening of the source radiation
\citep{CO95, GO00}. However, for simplicity, let us nonetheless base
our analysis on the spatially explicit form obtained by assuming a
canonical $\beta=1$ velocity law (\ref{betalaw}) within the finite
disk factor (\ref{fdcf}) for a simple spherical expansion. In the
zero sound-speed limit, the scaled equation of motion
(\ref{dimlessEoM}) can now be written in the form, \beq w' = -1 +
\omega^{2} (1-x) + fC_{c}\left (  { w' \over \dot{m}} \right
)^{\alpha} \, , \label{dimlesseom} \eeq where we have normalized the
line force in terms related to the ``point-star'' CAK model, with
$\dot{m} \equiv {\dot M}/{\dot M}_{\rm CAK} $, the ratio of the
mass-loss rate to the point-star CAK value.
Note then that for the non-rotating ($\omega=0$),
point-star ($f=1$) case of the classical CAK model, the critical
solution (with maximal mass loss) is given by $\dot{m}=1$ and
$w(x)=\alpha/(1-\alpha) \, x$. As noted in \S
\ref{subsec:CAKsoln}, this implies a CAK mass loss rate ${\dot M}
= {\dot M}_{\rm CAK}$ and a
velocity law, $v(r) = v_{\infty} \sqrt{1-\Rstar/r}$, with terminal
speed $ v_{\infty} = \sqrt{\alpha/(1-\alpha)} v_{\rm esc}$.

To analyze models with rotation, a particularly convenient case is
to take $\alpha=1/2$, for which the equation of motion
(\ref{dimlesseom}) (using eq. (\ref{Cc}) for $C_{c}$) becomes a
simple quadratic in $\sqrt{w'}$, \beq w' - 2f \sqrt{w'/\dot{m}} +
g(x) = 0 \, , \eeq where for convenience, we have defined a
rotationally reduced gravity as $g(x) \equiv 1 - \omega^{2} (1-x)$.
We can then solve for a shallow (--) and steep (+) acceleration
solution, \beq w'_{\pm}(x) = \frac{g(x) n(x)}{\dot{m}} \left [1 \pm
\sqrt{1-\dot{m}/n(x)} \right ]^{2} \, , \label{wppmsoln} \eeq with
the ``nozzle function'', \beq n(x) \equiv {f(x)^{2} \over g(x)} = {
f(x)^{2} \over 1 - \omega^{2} (1-x)} \, . \label{nozdef} \eeq

The significance of this nozzle function stems from its appearance
with the mass loss rate $\dot{m}$ within the square-root
discriminant(cf. Laval nozzle; \citet{Abbott1980}). In particular,
we can readily see that maintaining a numerically real flow
acceleration requires\footnote{Actually, this restriction really
stems from our CAK scaling of the line-force with $w'^{\alpha}$
(in this case $\sqrt{w'}$), which requires a strictly {\it
positive} acceleration, $w'>0$. But if we provide a backup scaling
for negative accelerations, then ``overloaded'' situations, for
which the square-root discriminant in eq. (\ref{wppmsoln}) becomes
negative, simply lead to an abrupt switch, a so-called ``kink''
\citep{CO96}, to a \emph{decelerating} solution. See \S
\ref{subsec:HydroSimSpecs} and \ref{sec:summary} for further
details.} a mass loss rate $\dot{m} \le \min[n(x)]$. As such, the
location of the global minimum of this function (the smallest
nozzle ``throat'') represents the {\it critical point} that sets
the maximal allowed value of the mass loss rate, $\dot{m}
=\min[n(x)]$, that is consistent with a monotonically accelerating
outflow.

\begin{figure}
\begin{center}
\plotone{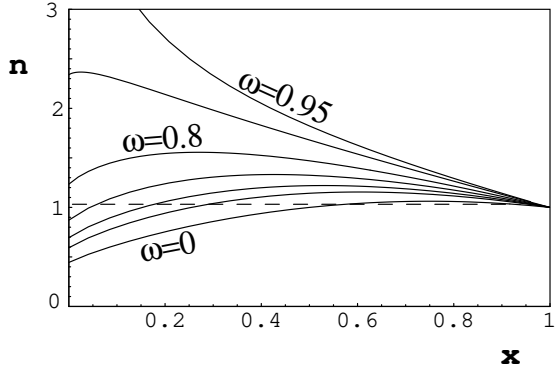} \caption{ Nozzle function
$n(x)$ plotted vs. scaled inverse radius $x = 1 - R_{\ast}/r$, for
various rotation rates $\omega =$ 0, 0.5, 0.6, 0.7, 0.8, 0.9,
0.95, ranging from lowermost to uppermost.
All curves use a $\beta=1$ velocity law in evaluating the
fdcf.
The horizontal dashed
line at unit value represents the nozzle function for the CAK
point-star model, with $n=f=g=1$. }
\label{fig2}
\end{center}
\end{figure}

Figure \ref{fig2} plots $n(x)$ vs. $x$ for various rotation rates
$\omega $, using a $\beta=1$ velocity law to obtain a spatially
explicit approximation to the fdcf. Note that for no or low
rotation (about $\omega < 0.75 $), the minimum of the nozzle
function is less than unity, and occurs at the stellar surface,
$x=0$. This allows the flow to transition to a {\it
super}-critical outflow directly  from the static  surface
boundary condition $w(0)=0$, following the steeper, plus (+) root
for the acceleration in equation (\ref{wppmsoln}), but with a mass
loss rate less than the point-star CAK value,
\beq \dot{m} = \dot{m_{0}} \equiv n(x=0) \equiv {f^{2} (x=0) \over
1-\omega^{2}} = {4/9 \over (1-\omega^{2}) } \, . \label{mdodef}
\eeq Note that the factor 4/9 in the numerator is just the
$\alpha=1/2$ value for the zero-rotation, finite-disk-corrected
mass loss scaling derived by FA and PPK, \beq \dot{m}_{\rm fd}
\equiv { \dot{M}_{\rm fd} \over \dot{M}_{\rm CAK} } =
f_{*}^{1/\alpha} = \frac{1}{(1 + \alpha)^{1/\alpha}} \, . \eeq

By contrast, for large rotation rates (about $\omega > 0.75$), this
nozzle minimum is unity, and occurs at large radii, $x=1$;
satisfying the static surface boundary condition now implies that the flow at all
finite radii should remain {\it sub}-critical, following the
shallower, minus (--) root for the acceleration in equation
(\ref{wppmsoln}), now with a mass loss rate just equal to the
point-star CAK value, $\dot{m}=1$.

This thus provides the basic explanation for the switch from steep to
shallow accelerations inferred by \citet{Cureetal05}.
\subsection{Nozzle Solutions for the Velocity Law in 1D Rotating Winds}
\label{subsec:numintsols}

\begin{figure*}
\begin{center}
\epsscale{1} \plottwo{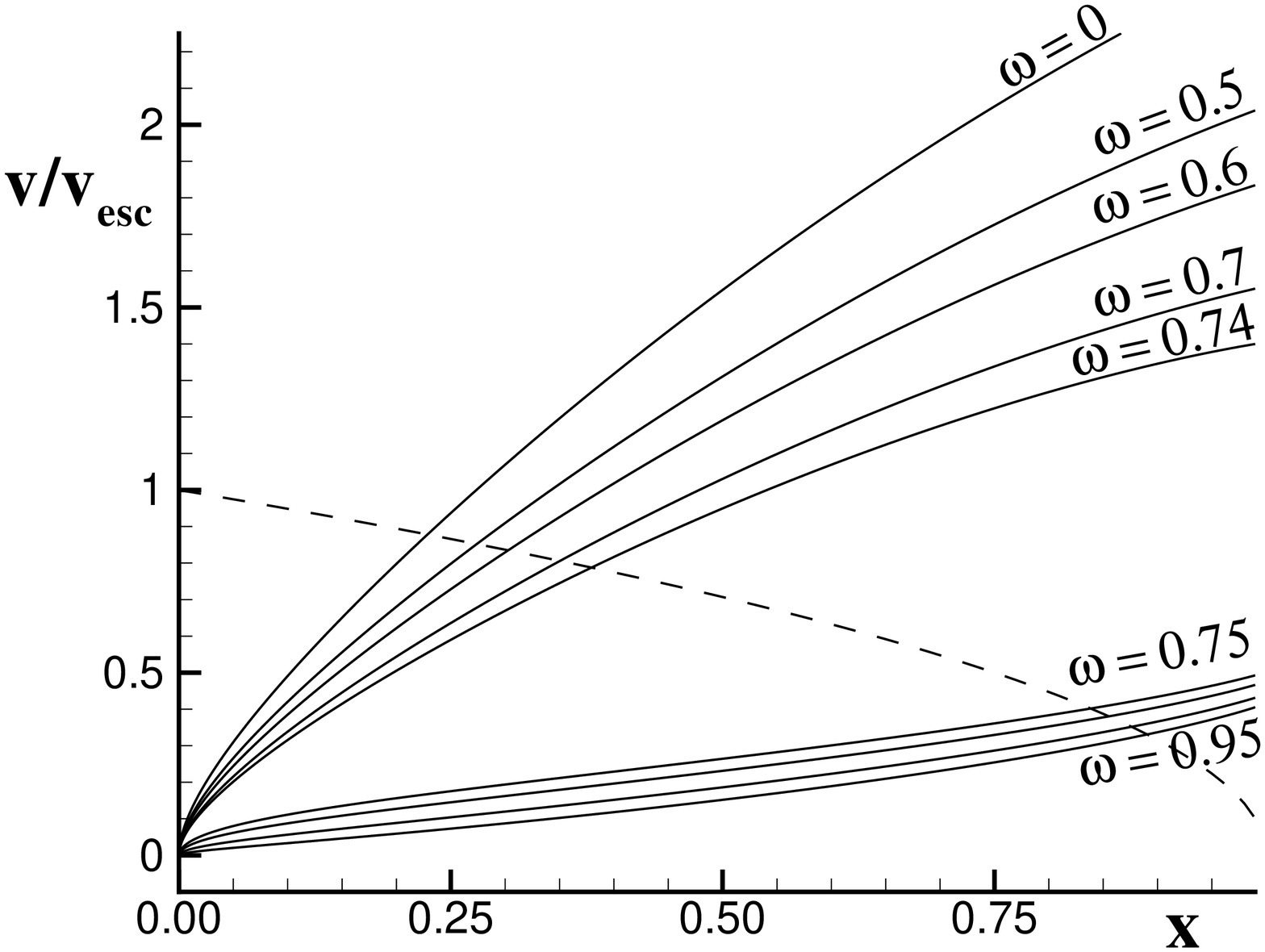}{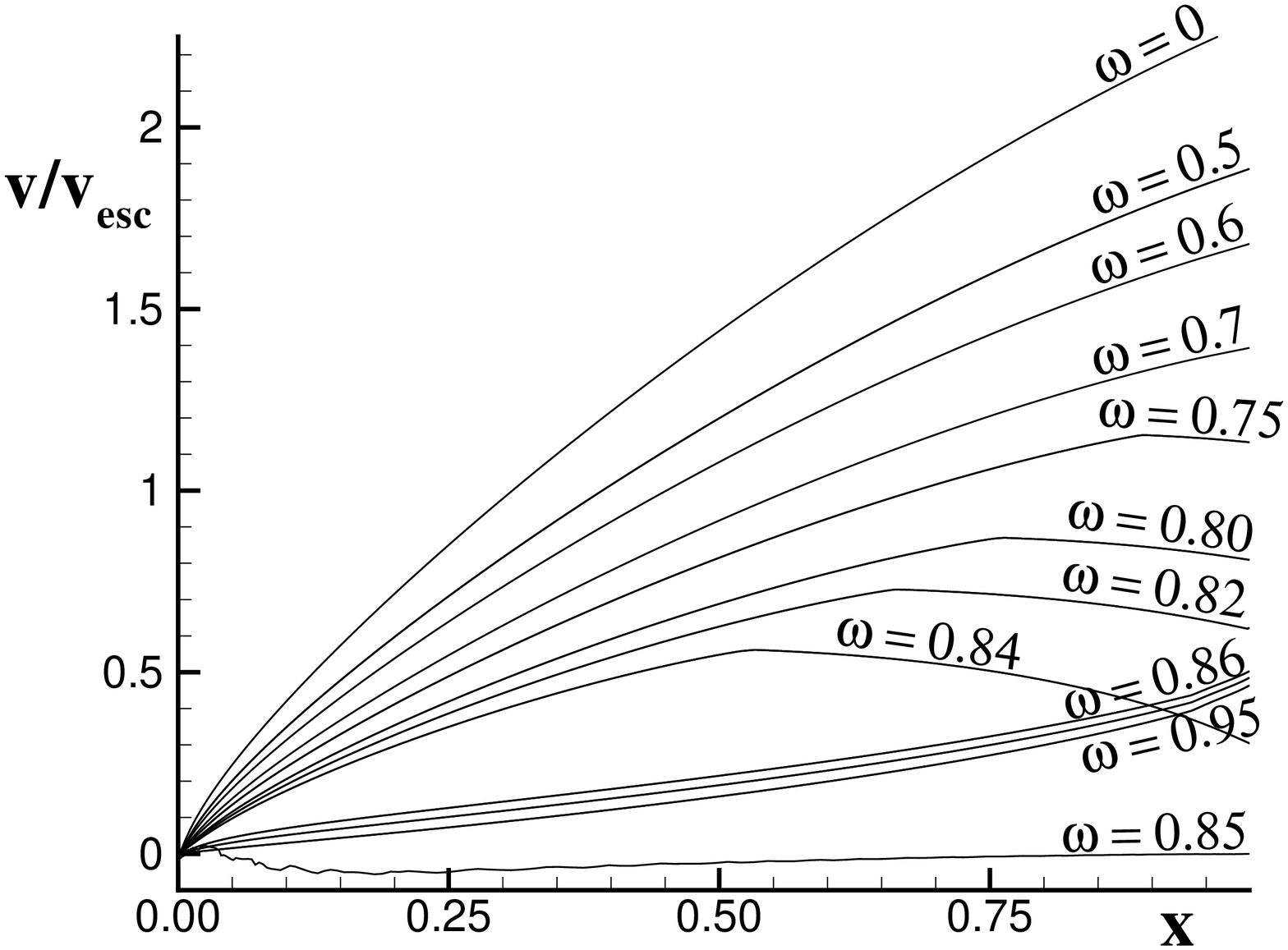} \caption{ (a) Flow speed
over escape speed, $v/v_{\rm esc} = \sqrt{w}$, plotted vs. scaled
inverse radius $x = 1 - R_{\ast}/r$, as derived from a nozzle
analysis, using a steep acceleration for no or modest rotation,
$\omega =$ 0, 0.5, 0.6, 0.7, 0.74 and shallow acceleration for
rapid, near-critical rotation, $\omega =$ 0.75, 0.8, 0.85, 0.9,
0.95. The dashed curve shows the escape speed as a function of $x$.
b) Same plot as in (a), but as found using asymptotic states of full
hydrodynamical simulations. Note again the steep, supercritical
accelerations for no or modest rotation, $\omega =$~0, 0.5, 0.7, and
shallow, subcritical accelerations for rapid, near-critical
rotation, $\omega =$0.86, 0.9, 0.95. However, note also the ``kink"
solutions present for $\omega =$~0.75, 0.8, 0.82, 0.84 and the
collapsed solution for $\omega =$~0.85.} \label{fig3}
\end{center}
\end{figure*}

The associated wind velocity laws can be obtained by simple
numerical integration of equation (\ref{wppmsoln}) from a static
boundary $w(0)=0$, following either the steep or shallow solution,
depending on whether the rotation rate is high enough to shift the
critical point (where $n(x)$ has its absolute minimum) from the
surface ($x=0$) to large radii ($x=1$). Figure \ref{fig3}a plots the
resulting velocity laws for selected slow vs. rapid rotation rates,
yielding respectively the steep vs. shallow types of flow solution.
The dashed curve in Figure \ref{fig3}a plots the escape speed
$v_{\rm esc}$ as a function of $x$, showing that these winds are
capable of escaping the star. The right panel compares results from
full dynamical simulations described below.

\begin{figure}
\begin{center}
\epsscale{1} \plotone{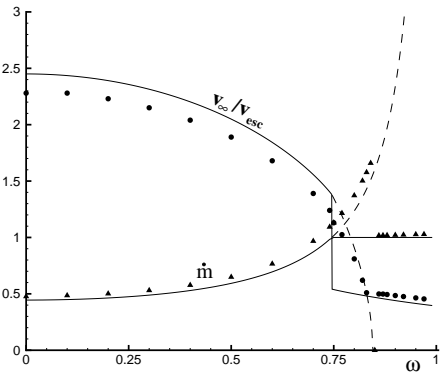} \caption{ Upper solid curve: Terminal
flow speed over escape speed, $v_{\infty}/v_{\rm esc} =
\sqrt{w(1)}$, plotted vs. rotation rate $\omega$, showing the shift
from fast to slow wind as rotation rate is increased past ca.
$\omega = 0.75$. Lower solid curve: Mass loss rate in units of
point-star CAK value, again plotted vs. rotation rate $\omega$,
showing the saturation at the CAK mass loss rate for rapid rotation,
$\omega > 0.75$. Dashed curves show the continued steady decrease in
$v_{\infty}/v_{\rm esc}$ and increase in $\dot{m}$ if the local
nozzle minimum at the stellar surface is used to set flow conditions
(\S\,\ref{subsec:numintsols}). The circles ($v_{\infty}/v_{\rm
esc}$) and triangles ($\dot{m}$) show corresponding results from
full hydrodynamical simulations. } \label{fig4}
\end{center}
\end{figure}

Figure \ref{fig4} illustrates the associated terminal speed and
mass loss rates for these solutions (solid lines), plotted as a
function of rotation rate $\omega$. As the rotation increases past
the threshold rate at $\omega \approx0.75$, the solid curves show
an abrupt shift from steep acceleration to shallow acceleration,
with the mass loss saturating to the point-star CAK value,
$\dot{m}=1$. The dashed curves show extrapolated results if the
local nozzle minimum at the surface is instead used to set flow
conditions; the mass loss in this case is set by the scaling
$\dot{m}_{0}$ in eq. (\ref{mdodef}), and the terminal speed is
derived by assuming a pure gravitational coasting for all radii
with $n(x) < \dot{m}_{0}$. The data points again compare
corresponding results for the full dynamical simulation, as
described further in the next sections.

%

\section{Specifications for Numerical Hydrodynamics Simulations}
\label{subsec:HydroSimSpecs}

The above nozzle analysis provides a helpful framework for
understanding the nature of flow solutions from a rotating wind
model, but is based on some key simplifications, e.g. neglect of gas
pressure (inclusion of which would lead to a wind that is not
super-critical directly from the static surface boundary) and an
approximate, spatially explicit form for the finite-disk correction
factor. Moreover, it implicitly assumes that the derived solutions
are the only relevant stable, attracting steady states for the
rotating wind outflow. To test the validity of these simplifications
and assumptions, let us now examine the \emph{time} evolution of
analogous 1D flow models, including also both the finite gas
pressure and the dynamically computed finite disk correction factor.
Our specific approach here is to use a numerical hydrodynamics code
to evolve a 1D time-dependent model of the equatorial plane a
line-driven stellar wind from a rotating star toward an equilibrium
steady-state for the resulting flow. The results of these
simulations can then be compared to those predicted by the
steady-state nozzle analysis presented above.

The numerical models presented here were computed using a
piecewise parabolic method \citep[PPM;][]{CW84} hydrodynamics code
called VH-1, originally developed at the University of Virginia
(J. Blondin). The basic code was modified for the present study to
include radiative driving terms,
solving the time-dependent equations for 1D spherical outflow
(\ref{hydro1}) and (\ref{hydro2}). The spatial mesh uses $n_{r} =
600$ radial zones extending from the base at the stellar surface,
$r_{1} = R_{\rm min} = R_{\ast}$, to a maximum $r_{600} =
R_{\rm max} = 100R_{\ast}$, with the zone spacing starting at
$\Delta r_{1} = 6.18 \times 10^{-5} \, R_{\ast}$ and then
increasing by 2.5\% per zone out to $r_{350} = 15R_{\ast}$, after
which it remains constant at $\Delta r = 0.333 \, R_{\ast} $ to the
outer boundary. Tests with double the grid resolution for selected
cases give similar results to the above standard values.

The parabolic method requires flow variables to be specified in a
phantom zone beyond each boundary. At the outer radius, we assume
supercritical outflow, with boundary conditions set by simple flow
\emph{extrapolation} assuming constant gradients. This is
justified because, as discussed in the Appendix, when finite
sound-speed terms are included, the critical point for even the
shallow acceleration solutions should be well within our assumed
outer boundary radius of $R_{\rm max} = 100 \Rstar$.

At the inner boundary, the velocity in the two radial zones below
$i_{\rm min}$ is set by constant-slope extrapolation, thus
allowing the base velocity to adjust to whatever is appropriate
for the overlying flow \citep{Owockietal94}. This usually
corresponds to a subsonic wind outflow, although inflow at up to
the sound speed is also allowed.
The base density is fixed at
$\rho_{0} = 8.709 \times 10^{-13}\;{\rm g}\;{\rm cm}^{-3}$, a
value chosen because, for the characteristic wind mass fluxes of
these models, it yields a steady base outflow that is moderately
subsonic. A lower-boundary density much smaller than this produces
a base outflow that is supersonic, and thus is unable to  adjust
properly to the mass flux appropriate to the overlying line-driven
wind. On the other hand, a much larger base density makes the
lower boundary too ``stiff'', leading to persistent oscillations
in the base velocity \citep{Owockietal94}.

These time-dependent simulations also require setting an
\emph{initial condition} for the density and velocity over the
entire spacial mesh at some starting time $t = 0$. For this we
generally use a standard, finite-disk-corrected CAK wind, computed
by relaxing a 1D, \emph{non-rotating} simulation to a steady
state; however, for selected models with moderately rapid
rotation, we also explore  using a slow-acceleration initial
condition (see \S ~\ref{subsec:varyIC}). From the assumed initial
condition, the models are advanced forward in time steps set to a
fixed fraction 0.25 of the Courant time.

Our version of VH-1 is set up to operate in CGS units, requiring
specification of physical values for the basic parameters for both
the star (e.g., mass, radius, luminosity) and wind (e.g., CAK $k$,
$\alpha$, and $\delta$). Building upon our earlier studies of Be
stars, the specific parameters chosen here are for a main sequence B
star with mass $\Mstar = 7.5M_{\odot}$, radius $\Rstar =
4R_{\odot}$, luminosity $\Lstar = 2310 L_{\odot}$, and temperature $
T = 2\times 10^{4}$~K; but we have also explored models with
parameters appropriate for supergiant B[e] stars. These stellar
parameters imply an Eddington parameter of $\Gamma_{e} = 0.008$, an
isothermal sound speed of $a = 16.6$~km/s, and escape and critical
speeds of $v_{\rm esc} = 845$~km/s and $v_{\rm crit} =597$~km/s. We
also assume a CAK power-law index of $\alpha = 1/2$ and cumulative
line-strength parameter $\overline{Q} = 1533$ \citep{Gayley95}. The
value of $\delta$ has been set to zero in all simulations.

In any case, for a given choice of the CAK power-law index $\alpha$,
we find that results are largely independent of the specific
physical parameters when cast in appropriately scaled units,
normalizing for example radius by the stellar radius $\Rstar$,
velocity by the wind terminal velocity $v_{\infty}$ (which in turn
scales with the stellar surface escape speed), time by the
characteristic flow time $\Rstar/v_{\infty}$, and mass loss rate in
terms of the classical (point-star) CAK value given in eqn.
(\ref{MdotCAK}). To faciliate comparison with the analytic nozzle
analysis in \S\S 2-3, we again choose $\alpha=1/2$, and plot all
simulation results using the above scalings.

We further note that essentially all the key VH-1 results reported
here were very well reproduced by a completely independent, simple
dimensionless hydrodynamics code developed by one of us
\citep[AF;][]{FN06}.

Finally, in such time-dependent simulations of line-driven winds,
one must also supply a generalized scaling for the line-force that
applies in the case of non-monotonic flow acceleration.
In general this requires taking into account \emph{non-local}
couplings of the line-transfer \citep[see, e.g.,][]{FN06}, but we
wish here to retain the substantial advantages of using a purely
local form for the line-driving. Noting that a negative velocity
gradient implies a prior line resonance that shadows radial photons
from the star, a lower limit would be just to set $g_{\rm line} = 0$
whenever $dv/dr < 0$. On the other hand, since forward scattering
can substantially weaken any such shadowing by a prior resonance, an
upper limit would be to compute the local line force using the
absolute value of the velocity gradient, $g_{\rm line} \propto
|dv/dr|^{\alpha}$. As a simple compromise between these two
extremes, we choose here a scaling that truncates the radial
velocity gradient to zero whenever it is negative, i.e.  $dv/dr
\rightarrow \max(dv/dr,0)$. For a point-star model with radially
streaming radiation, this would give a zero line-force (since $dv/dr
<0$), but when one accounts for the lateral expansion $v/r$ within
the finite-disk correction factor, it leads to a line force in which
the usual dependence on radial velocity gradient is replaced by a
dependence on the expansion gradient, \beq g_{\rm line} \propto
(v/r)^{\alpha} \, . \eeq This leads to a line acceleration that is
intermediate between the underestimate and overestimate of the two
more extreme scalings.

\section{Comparison with Time-Dependent Hydrodynamical Simulations}
\label{sec:CompTDSims}

\subsection{Asymptotic Steady-States of Time-Dependent Simulations}
\label{subsec:AsympState}

In our basic parameter study, each simulation is run using the same initial
condition, spatial mesh, boundary conditions, base density, etc.,
with the only variation being the rotation rate $\omega$, which is
set to specific values ranging from 0.1 to 0.97.
%
With only one exception (for the $\omega=0.85$ case, which turns
to be a rather pathological value; see \S ~\ref{subsec:TimRelax}),
all simulations asymptotically relax to a well-defined steady
state. Moreover, for both moderate rotation ($\omega < 0.75$) and
high rotation ($\omega > 0.85$), these asymptotic states agree
remarkably well with the predictions of the above nozzle analysis.
Figure \ref{fig3}b shows the velocity laws for these final states,
scaled in the same form used in figure 3a for the nozzle-analysis
results.
Note that both figures 3a and 3b show a steep acceleration for
no or modest rotation ($\omega =$~0--0.7),
and shallow acceleration for rapid, near-critical rotation
($\omega =$~0.86, 0.9 and 0.95).

However, for the moderately high rotation-rate cases $\omega
=$~0.75, 0.8, 0.82 \,and 0.84, note also the appearance of a new
class of \emph{kink} solutions, characterized by an abrupt shift
to a decelerating or ``coasting" flow beyond a well-defined ``kink
radius'' $r_{\rm kink}$. These kink solutions thus represent a
kind of intermediary final state of the time-dependent simulations
in the parameter ranges $ 0.75 < \omega < 0.85$, effectively
smoothing the abrupt jump from fast to slow acceleration solutions
expected from steady-state analyses. The formation of such kinks,
and their underlying physical cause, are discussed further in the
section below (\S ~\ref{subsec:TimRelax}) on time evolution.

Figure \ref{fig4} shows that the fully dynamical results for the
scaled ratio of right boundary speed (circles) and CAK scaled mass
loss rate (triangles) are generally in good agreement with the
predictions of the simple nozzle analysis (solid curves),  with
the modest, ca. 10\% differences likely attributable to inclusion
in the simulations of a small but finite sound speed
\citep{OuD04}. However, for rotation rates $0.75 < \omega < 0.85$,
the dynamical results tend to follow the dashed curves of the
extended nozzle analysis, representing extended fast solutions
rather than the abrupt shift to slow solutions indicated by the
solid curves. As discussed above, this range of rotation rates is
characterized by kink solutions. To understand better this
development of fast vs. slow vs. kink solutions, let us now
examine the time evolution of the simulations toward asymptotic
states.


\subsection{Time Relaxation of 1D Rotation Models}
\label{subsec:TimRelax}

%

\begin{figure*}
\begin{center}
\plottwo{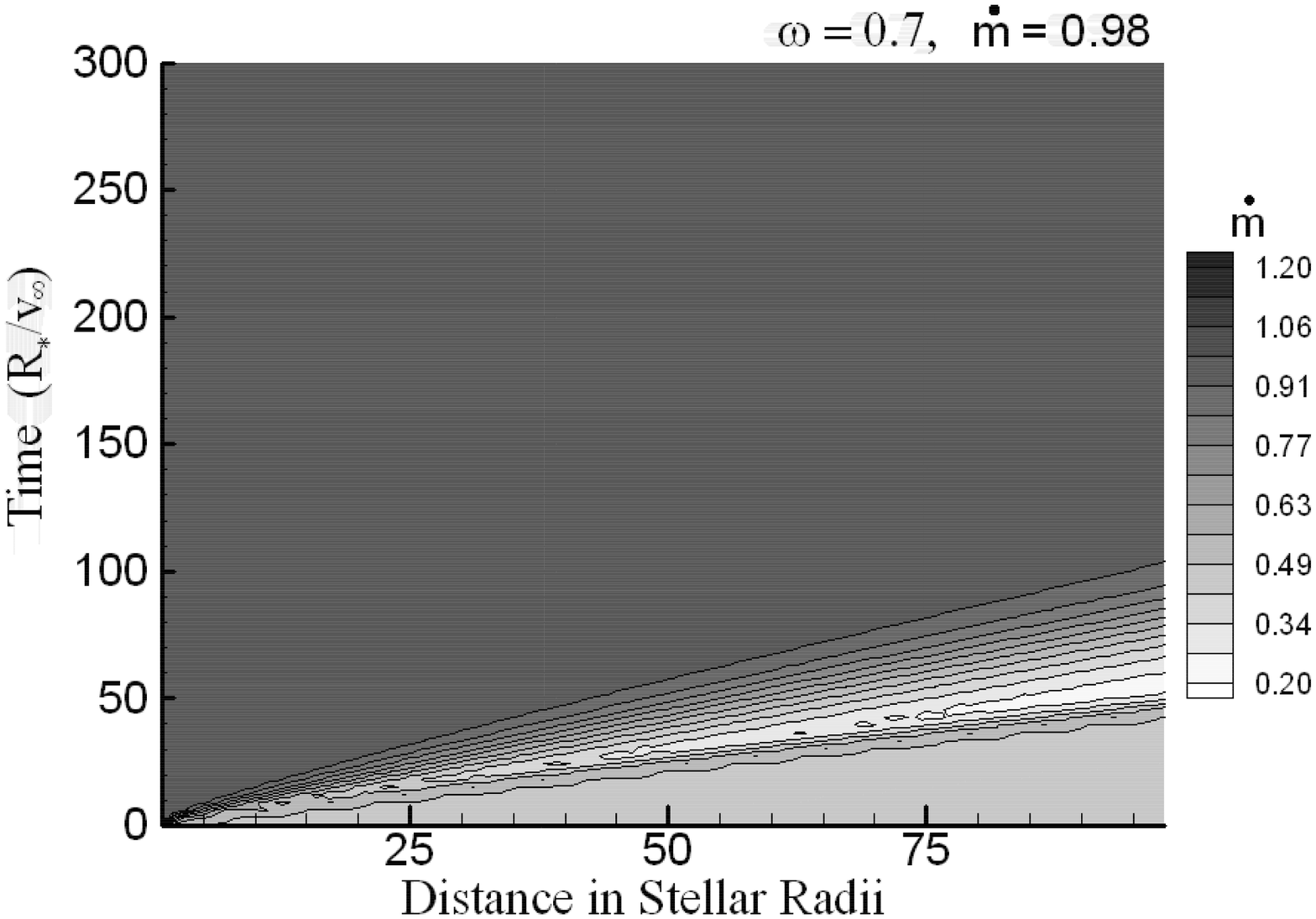}{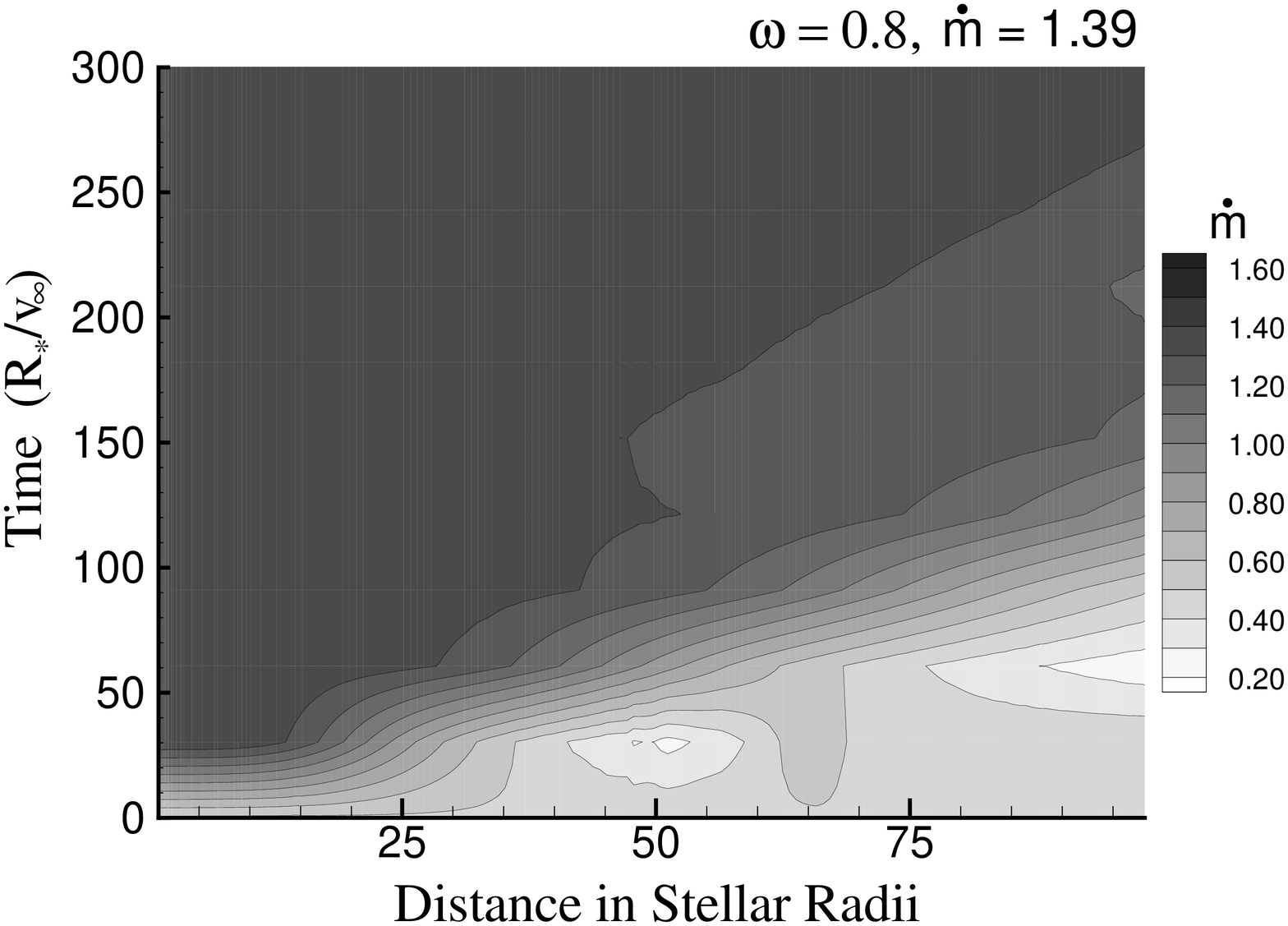}
\end{center}
\begin{center}
\plottwo{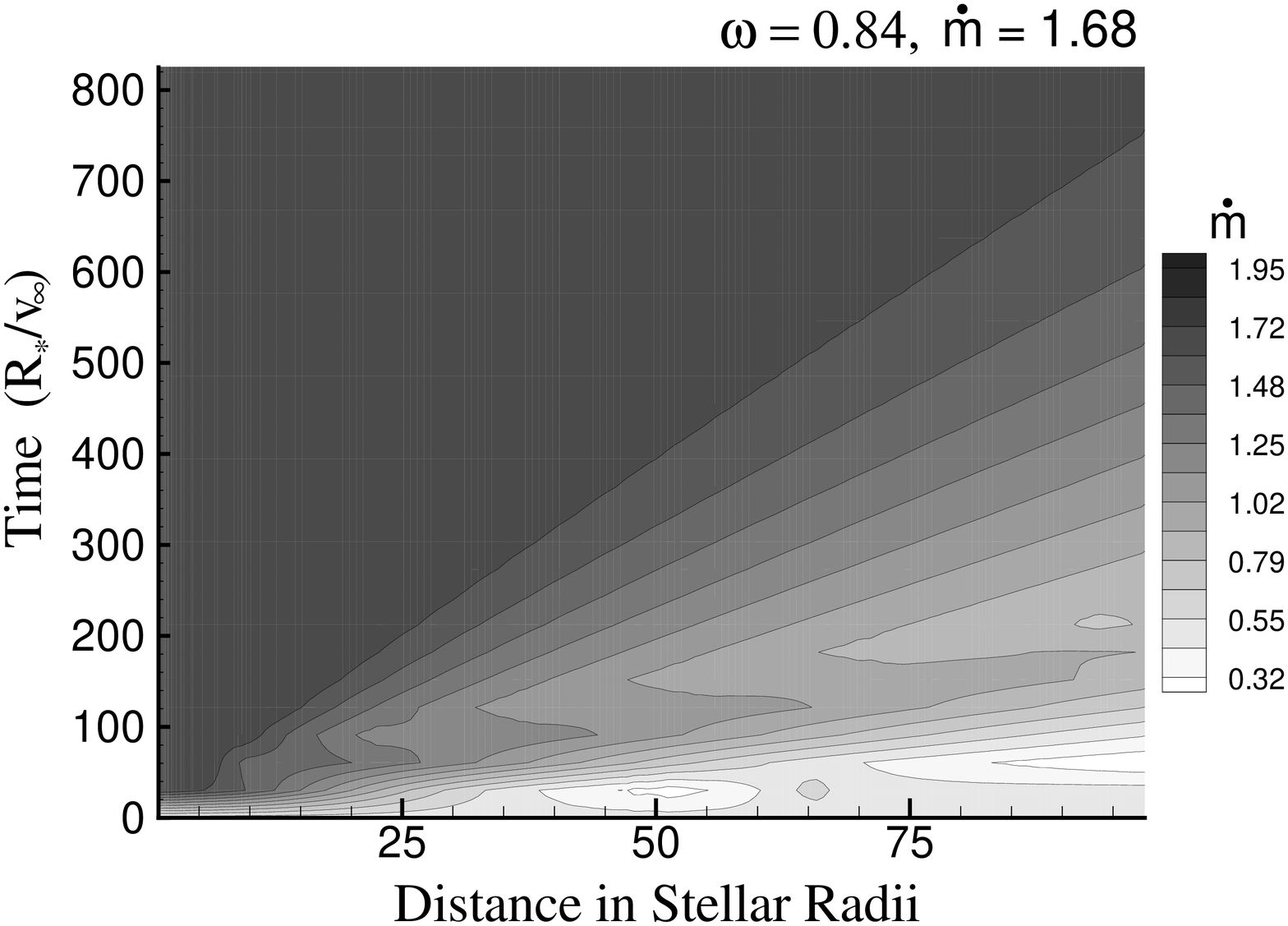}{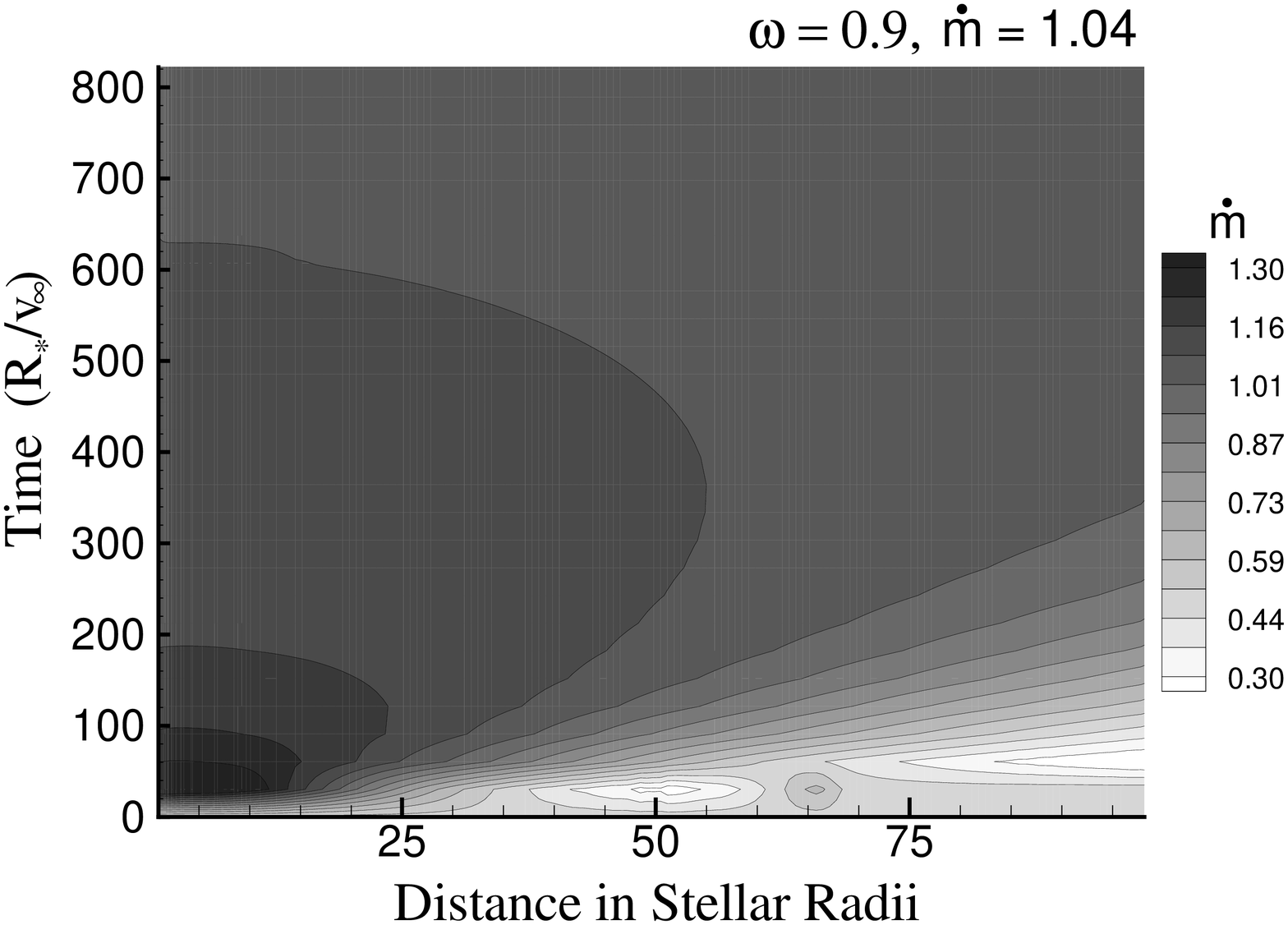} \caption{ Gray-scale
plots showing time evolution of mass-loss rate in units of the
point-star CAK value. The corresponding rotation rate and final
mass-loss rate are given above each plot. Time is in units of the
flow time $t = R_{\ast}/v_{\infty}$. Note in each the eventual
constancy of the mass-loss rate with both radius and time,
indicating relaxation to a steady-state solution. Note that the
ranges for both gray-scale and time differ for each panel. }
\label{fig6}
\end{center}
\end{figure*}

For the specific rotation rates $\omega =$ 0.7, 0.8, 0.84, $\&$
0.90, which span the parameter range between fast and slow
acceleration solutions, Figure \ref{fig6} uses gray-scale plots of
the mass-loss rate (in units of the point-star CAK value) to
illustrate the time relaxation from the CAK initial condition (set
to a non-rotating, finite-disk corrected CAK model) to an asymptotic
steady-state. Time is given in units of the flow time $t =
R_{\ast}/v_{\infty}$, and the final values of the scaled mass-loss
rate are indicated above the associated plot. Note that there are
distinct differences in the time evolution of each model, with the
more rapid rotation cases characterized by a longer relaxation time,
but the temporal and spatial constancy of the final mass loss rates
illustrate the steady nature of the asymptotic solutions.

\begin{figure*}
\begin{center}
\plottwo{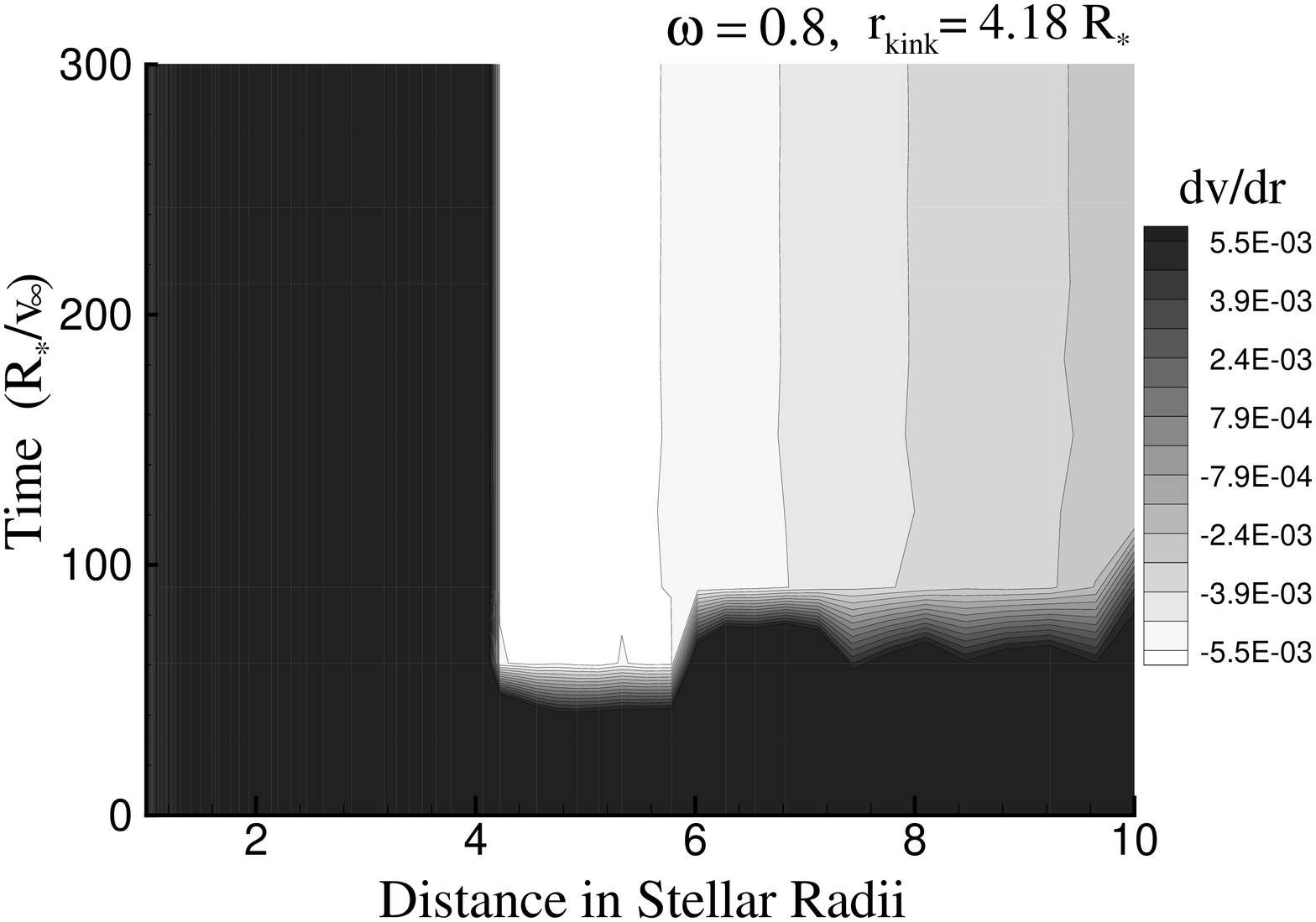}{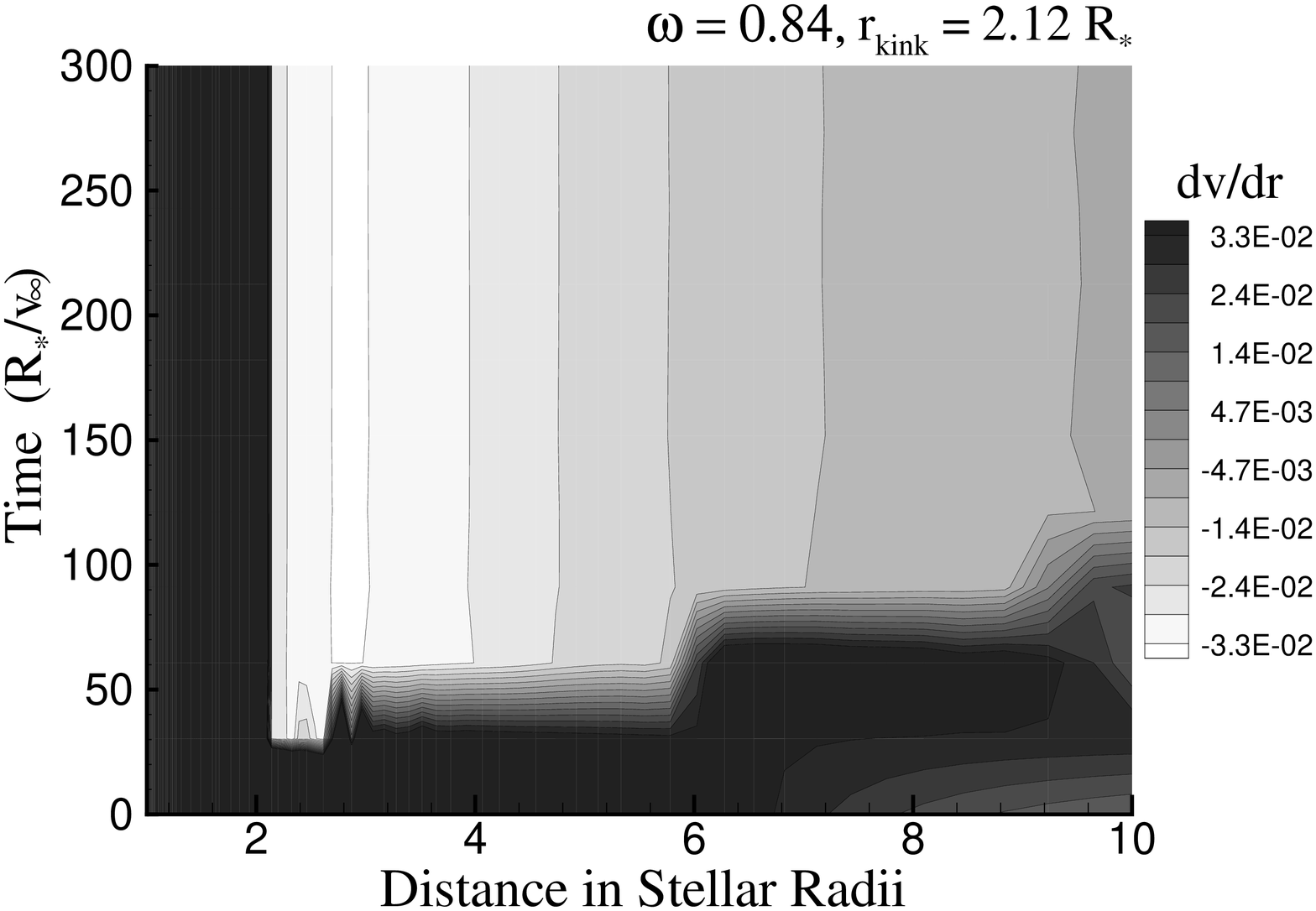} \caption{ Gray-scale
plots showing time evolution of the scaled velocity gradient in cgs
units. The corresponding rotation rate and kink location $r_{\rm
kink}$ are given above each plot. } \label{fig7}
\end{center}
\end{figure*}

Despite this temporal and radial constancy in the mass
flux, the asymptotic states of the models with $\omega =$~0.8 and
$0.84$ include a kink, or abrupt discontinuity in their velocity
gradient. To illustrate the formation of these kinks, the
gray-scales in figure \ref{fig7} show the evolution of the scaled
velocity gradient. In both cases, the abrupt switch from a steep
acceleration solution to a decelerating solution is apparent from
the sharp transition from positive to negative velocity gradient,
with the kink radius, $r_{\rm kink}$, indicated above each plot.
Since, despite the radial discontinuity, the velocity gradient
contours all become constant in time, we see clearly that the kink
solutions are indeed perfectly valid steady-states.

Note, however, that the kink radius shifts from  $r_{\rm
kink} = 4.18 R_{\ast}$ for the $\omega = 0.8$ model to $r_{\rm kink}
= 2.12 R_{\ast}$ for the more rapid-rotation, $\omega = 0.84$ model.
From the velocity law plots in figure \ref{fig3}b, this also implies
that the more rapid rotating case has a more extended deceleration
region, and thus ends up with a much lower final speed. We can thus
anticipate that a somewhat faster rotation rate should give an even
lower kink radius, with the more extended decelerating region
leading to an even lower final speed, or perhaps even to a flow
stagnation (zero velocity) at a finite radius. In particular, note
from figure \ref{fig4} that the dashed extrapolation curve suggests
that the onset of such flow stagnation should occur near a rotation
rate of $\omega \approx 0.85$.

\begin{figure*}
\begin{center}
\plottwo{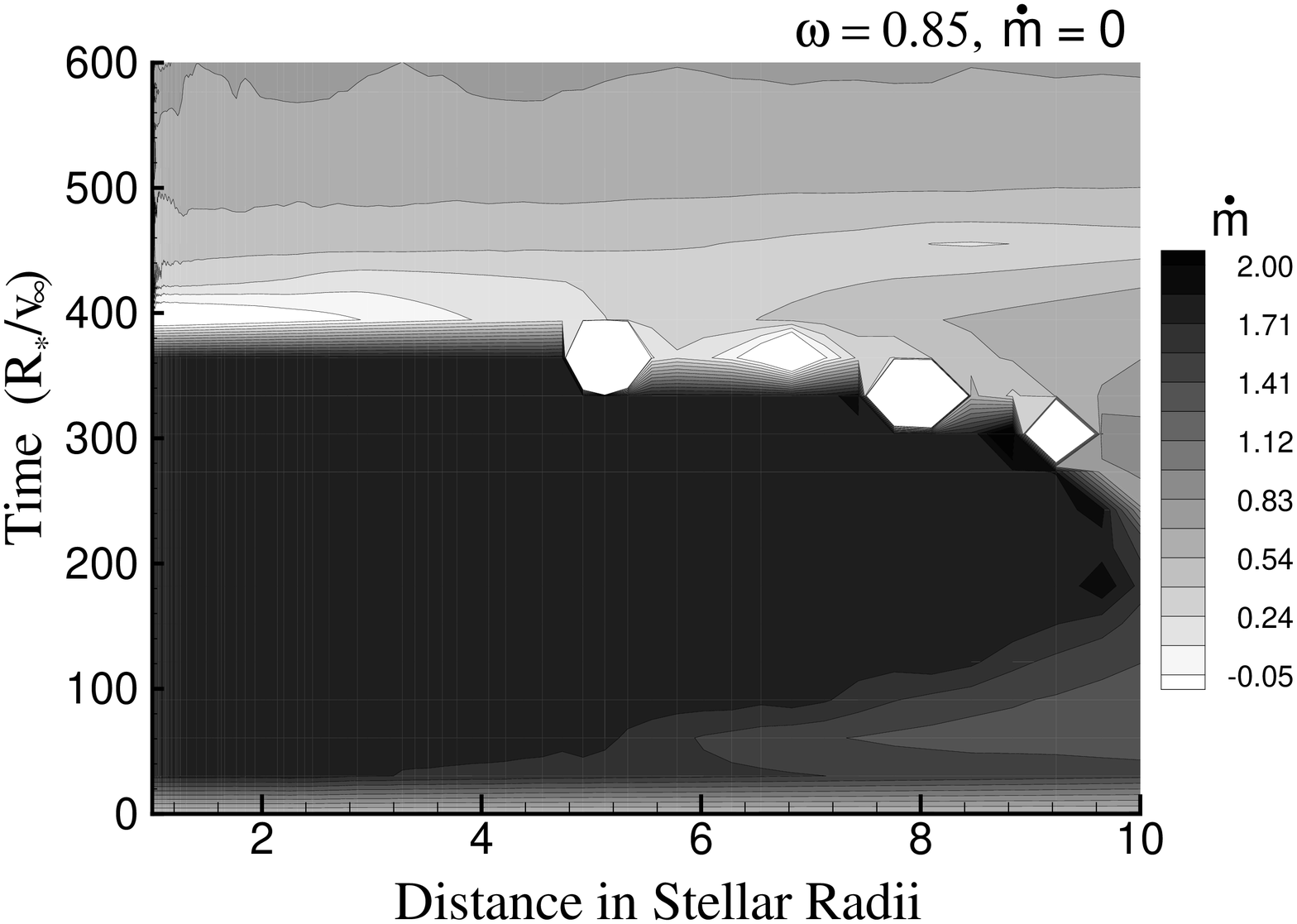}{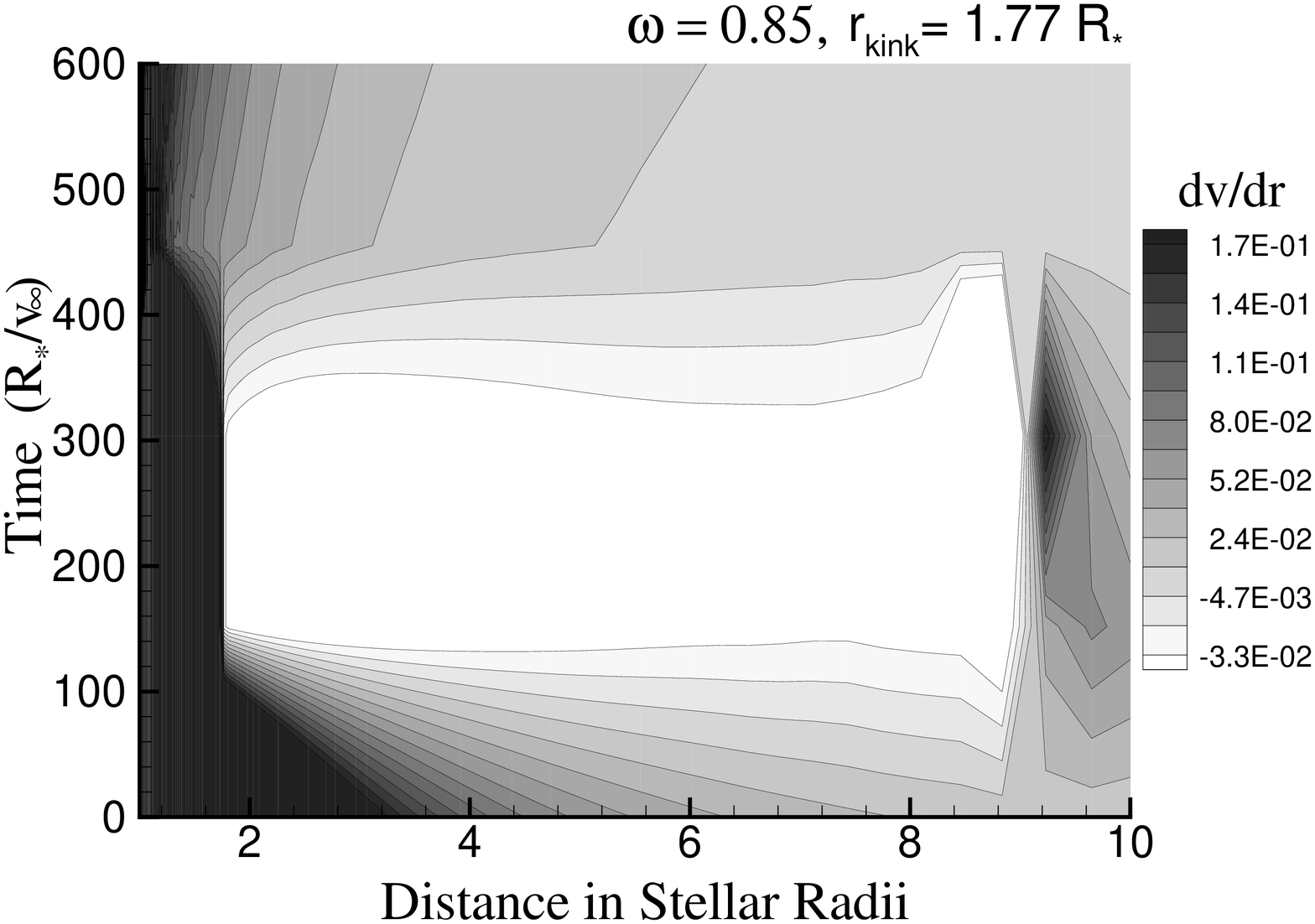} \caption{ (a) Gray-scale
plot of the local mass-loss rate in units of the point-star CAK
value for $\omega = 0.85$, showing the collapse of the solution.
Note the pile-up of material at a distance of $r \approx
10R_{\ast}$, and the eventual stagnation of the wind and
reaccreation of material back onto the stellar surface. (b)
Gray-scale plot of the scaled velocity gradient in cgs units for
$\omega = 0.85$. In the time interval between $t = 150$ and 300 flow
times, a kink in the velocity gradient is clearly visible, with the
kink location at $r_{\rm kink} \approx 1.77 R_{\ast}$. After $t
\approx$ 300, this kink solution becomes overloaded and unstable,
eventually collapsing.} \label{fig8}
\end{center}
\end{figure*}

In fact, our numerical simulations do show that models near this
rate have a quite pathological behavior. This is illustrated in
figure \ref{fig8}, which presents gray-scale plots of the
time-evolution for (a) the mass-loss rate and (b) the scaled
velocity gradient in this $\omega = 0.85$ case. An overloading of
the wind and eventual collapse of the time-dependent solution are
shown in these figures. In the first 300 flow times, there is an
initial kink formation, but at its lower radius of $r_{\rm kink} =
1.77 R_{\ast}$
the kink outflow speed is quite low, $v_{\rm kink} \approx
370$~km/s, well below the local escape speed $v_{\rm esc} (r_{\rm
kink}) \approx 640$~km/s. With the reduced line-driving in the
decelerating region, the wind outflow now \emph{stagnates} at a
finite radius, $r_{\rm stag} \approx 10 R_{\ast}$.
There material accumulates until it is eventually pulled back by
the stellar gravity into a reaccretion onto the star,
effectively quenching both the kink and base outflow.

For this particular case, the outflow never fully recovers from this
quenching, but for only slightly more rapid rotation,
$\omega \gtwig 0.86$, any flow stagnation occurs relatively close to
the star, with a correspondingly faster and less massive reaccretion,
followed by a recovery to a \emph{slow} acceleration solution, as
indicated in figures \ref{fig3} and \ref{fig4}.
In the very rapid, near-critical rotation case $\omega =$~0.90,
no kink or flow stagnation forms, and the solution relaxes more
directly to the slow solution, as illustrated in the lower right panel
of figure \ref{fig6}.

Finally, table 1 compares the radius of the kinks found in our
numerical simulations with the radius at which the nozzle analysis
indicates a steep acceleration solution can no longer be maintained.
For cases in which the mass flux set at the base $\dot{m} =
\dot{m}_{0} = n(x=0)$ is above the CAK value, i.e. $\dot{m} =
n(x=0)> 1$ , this occurs at a radius $x_{k}$ where a declining
nozzle function falls back to $n(x_{k}) = \dot{m}$. The
corresponding radius $r_{\rm kink, nozzle}$ agrees quite well with
the kink locations found from the hydrodynamical simulations,
$r_{\rm kink, hydro}$. Note also that the requirement $n(x=0) =
(4/9)/(1-\omega^{2}) > 1$ implies rotation rates of $\omega >
\sqrt{5}/3 \approx 0.745$, representing the onset for either
possible kink solutions or a switch to a slow acceleration solution.

\begin{deluxetable}{ccc}

\tablecaption{Comparison of Kink Locations }
\tablenum{1}

\tablehead{\colhead{$\omega$} & \colhead{$r_{\rm kink, hydro}$}
& \colhead{$r_{{\rm kink, nozzle}}$} \\
\colhead{} & \colhead{($R_{\ast}$)} & \colhead{($R_{\ast}$)} }

\startdata
0.80 & 4.18 & 4.54 \\
0.82 & 2.95 & 3.03 \\
0.83 & 2.52 & 2.54 \\
0.84 & 2.12 & 2.16 \\
0.85 & 1.77 & 1.86 \\
\enddata

\tablecomments{$r_{\rm kink, hydro}$ and $r_{\rm kink, nozzle}$
are the kink locations found respectively from numerical
hydrodynamical models and from the semi-analytic nozzle analysis.
}

\end{deluxetable}

\subsection{Results for Slow Acceleration Initial Condition}
\label{subsec:varyIC}

A central finding of the dynamical simulations is that moderately
fast rotation models $ 0.75  < \omega < 0.85$ form
fast-acceleration kink solutions instead of the slow acceleration
solutions predicted from steady-state analyses. But since the
non-linear character of the flow equations allows more than one
solution, this raises the question of whether slow acceleration
solutions in this regime might also be stable attractors, perhaps
for initial conditions that are closer to their slower outflow
form than the fast, non-rotating model used for the initial
conditions in the simulation models discussed above.
For each of the specific rotation cases in this transitional
range, $\omega =$~0.75, 0.80, 0.82, and 0.84, we thus recompute
simulations that instead use an initial condition set to the slow
acceleration steady-state found for the faster, near-critical
rotation case $\omega=$~0.9.

For the $\omega=$~0.75 case, we find that the model again relaxes
to a fast solution with an outer-wind kink at $r_{\rm kink}
\approx 10 \Rstar$. Steep acceleration solutions are also
recovered in all the slower rotation models as well.

\begin{figure}
\begin{center}
\epsscale{1} \plotone{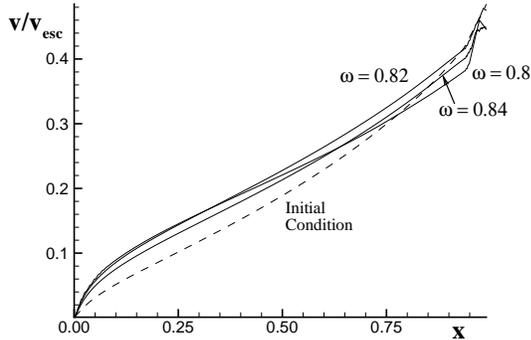} \caption{ Final wind velocity law
(flow speed over escape speed, $v/v_{\rm esc}$, plotted vs. scaled
inverse radius $x$) for $\omega = 0.8$, 0.82, and 0.84 simulations
that used the $\omega = 0.9$ final state (dashed line) as an initial
condition. The solid lines now approximate the expected slow
acceleration solution, except for an unexplained uptick in velocity
gradient near the right boundary, $x > 0.9$. } \label{fig10}
\end{center}
\end{figure}

However, as shown in figure \ref{fig10}, for the $\omega$ =~0.80,
0.82, and 0.84 cases the final states now do approach slow
acceleration solutions, apart from a persistent peculiar upward
kink near the outer boundary (i.e. for $x>0.9$ in figure
\ref{fig10}). Such models also show persistent small-scale
fluctuations with an amplitude of ca. 10\% in the mass flux,
apparently reflecting some difficulty for the numerical solution
to relax to a subcritical flow solution in these cases. The upward
kink may  stem from using a super-critical outflow boundary
condition for this slow acceleration solution, which does not
become supercritical until far from the star.

Finally, even for this slow acceleration initial condition, the
peculiar case $\omega$=0.85 still forms an overloaded condition with
flow stagnation and reaccretion, after which it never fully recovers a
steady outflow result.

\begin{figure}
\begin{center}
\epsscale{1} \plotone{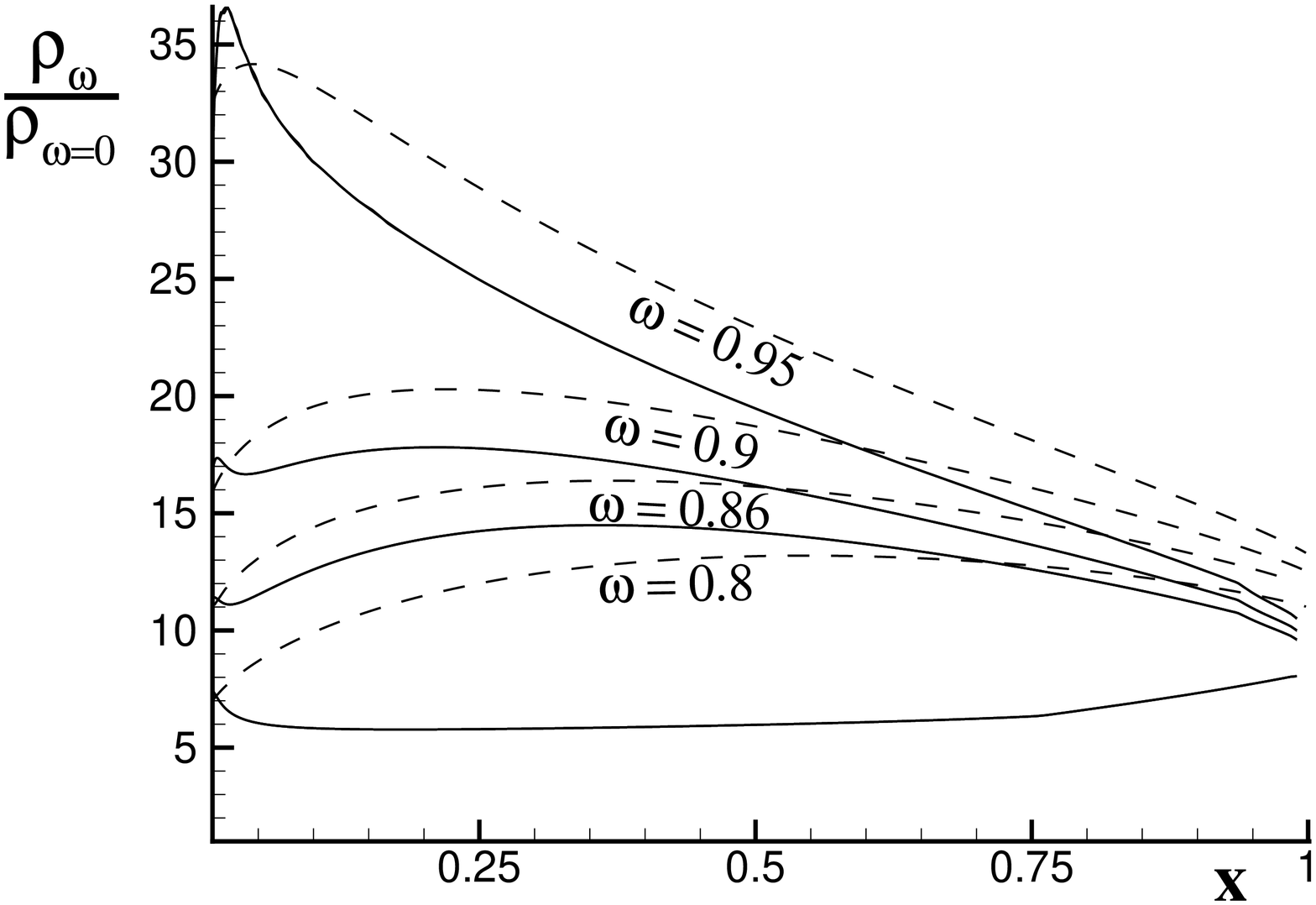} \caption{ Density
enhancement of slow wind solutions with rapid rotation rates,
$\omega=$ 0.8, 0.86, 0.9, and 0.95, relative to a non-rotating
wind with the same wind parameters. Dashed curves are for analytic
results, while solid curves are those obtained from numerical
simulations using VH-1.} \label{fig9}
\end{center}
\end{figure}
\subsection{1D Results for Equatorial Density}
\label{subsec:eqDen}

These results also allow us to identify the radial variation of
the relative density enhancement in the slow equatorial wind of a
rotating star, compared to the non-rotating solution that applies
to the polar wind. We are ignoring here gravity darkening and
oblateness effects, as well as any ``bi-stability'' in the
line-driving parameters between the polar and equatorial wind.
From the analysis of \S\,\ref{sec:SSsolns}, the relative density
enhancements are given by the ratios of the quantity
$\dot{m}/\sqrt{w}$ between the rotating and non-rotating models.

The dashed curves in figure \ref{fig9} show the spatial variation of
this density enhancement for rotating models with $\omega=$ 0.8,
0.86, 0.9, and 0.95. Note that the enhancements are a few factors of
ten, not insignificant, but not sufficient to reproduce the inferred
densities of B[e] disks, which are factors of order $10^{4}$ or more
denser than a typical polar wind outflow \citep{Zickgrafetal85,
KrausMiro06}. Inclusion of bi-stability effects could give about
another factor of a few by increasing the equatorial mass loss
around the cooler, gravity-darkened equator, thus yielding an
overall enhancement of order $10^{2}$ \citep{LP91, Pelupetal00}.
Significantly higher enhancement would require an
unrealistically low $\alpha$ to increase further the equatorial
mass flux, and/or assuming an equatorial surface rotation within a
sound speed of the critical (orbital) speed. In the latter case,
minor disturbances (e.g. pulsations) in the stellar envelope or
photosphere could instead eject material into an orbiting,
Keplerian disk \citep{Lee91, Owocki05}, obviating the need to
invoke any central role for radiatively driven outflow solutions.

\section{Summary and Conclusions} \label{sec:summary}

Using an analytic approach combined with numerical hydrodynamic
simulations, we investigate the reasons for the switch from a steep
to shallow acceleration in 1D line-driven stellar wind models as
stellar rotation rates are increased beyond a threshold value of
$\omega \approx 0.745$. The results indicate that the cause of this
switch is the overloading of the base mass-loss rate beyond the
point-star CAK value. The latter represents the maximal allowed mass
loss for which there can be a \emph{monotonically} accelerating flow
speed throughout the entire wind. Furthermore, the finite-disk
correction-factor (fdcf) reduces the driving effectiveness near the
stellar surface, and thus reduces the maximal mass loss that can be
initiated there. This reduction allows the outer wind to maintain a
positive acceleration even as other effects (e.g. centrifugal
reduction in effective gravity near the surface) allow for an
increase of the base mass loss from its fdcf value.

This problem of wind overloading at large rotation rates was first
noticed by \citet{FA}, who found indeed that beyond some threshold
rotation rate, the super-critical, steep acceleration solutions
they were deriving could not be followed beyond some finite
radius.
The contributions of M. Cur$\acute{e}$ and collaborators
have since shown that this termination can be avoided by switching
to a shallow acceleration solution.
As rotation increases beyond
the threshold value of $\omega \approx 0.745$, the base mass-loss
becomes greater than the point-star CAK value, and so the only
\emph{globally accelerating} solution possible is a shallow one
with a subcritical outflow.

However, an important lesson learned from our numerical
simulations is that the flow \emph{does not} necessarily follow
the solution with a globally monotonic acceleration.
We see that
the steady 1D solutions for rotating winds actually fall into
\emph{four} domains.
First, for no or low rotation rates ($\omega
< 0.745$), the minimum of the nozzle function is less than unity
and occurs at the stellar surface (x = 0).
This allows the flow to
transition to a super-critical outflow directly  from the static
surface boundary and follow the steep acceleration solution with a
mass-loss rate that is less than the point-star CAK value.

Next, there exits a ``gray zone" for rotation rates between $\omega
= 0.745$ and $\sim$ 0.86 where two different solutions are possible.
If we restrict ourselves to a strictly positive acceleration, $w' >
0$, then as rotation increases beyond $\omega \approx 0.745$ the
base mass loss exceeds the point-star CAK value, and so the only
globally accelerating solution possible is one with a shallow
acceleration and a mass-loss rate that saturates to the point-star
CAK value. On the other hand, if we provide a backup scaling for
negative accelerations and investigate cases where multiple local
minima appear in the nozzle function, then overloaded situations
where the square root term in equation (\ref{wppmsoln}) becomes
negative lead to a ``kink", and thus a negative acceleration or
coasting solution. Moreover, for these new kink solutions, the
multiple minima of the nozzle function now give rise to mass-loss
rates that exceed the point-star CAK value.

This ``gray zone" of kink solutions can furthermore be divided into
two domains. For rotation rates between $\omega \approx 0.745$ and
$\sim 0.85$, the overloading of the wind is not too severe. The
coasting solution can thus still reach large distances with a finite
speed, and so, these cases are able to approach a perfectly steady,
time-independent state, despite the presence of the discontinuity in
slope at the kink. On the other hand, for rotation rates of $\omega
= 0.85$, the wind becomes strongly overloaded and eventually
stagnates at some finite radius. After enough material has piled up
at this radius, it will fall back as a ``reaccreation front" toward
the star. If the boundary conditions are such that this material is
allowed to fall through the lower boundary, then it effectively
disappears from the model allowing the simulation to re-establish a
wind outflow with a slow acceleration solution (as in the $\omega =
0.86$ case).

For rotation rates greater than $\omega \approx 0.86$, the global
minimum of the nozzle function is unity and occurs at large radii
($x = 1$). In order to satisfy the static surface boundary
condition, the flow at all finite radii should be subcritical (but
still supersonic), and so the only solution possible is the shallow
($-$) acceleration solution, with a mass-loss rate that has
saturated to the point-star CAK value.

It is also worth noting that since the non-linear nature of the flow
equations allow more than one solution, slow acceleration solutions
in the $0.75 < \omega < 0.85$ regime are also possible via a change
in initial conditions. By initializing the wind using the slow
acceleration steady-state found for the $\omega = 0.9$ case, it is
possible to achieve final states that do approach the expected slow
acceleration solution (apart from a peculiar upward kink near the
outer boundary, see fig. \ref{fig10}).

One should also add here a few words regarding the validity of the
Sobolev approximation for the slow acceleration solutions. Recall
that for the usual case where the broadening of the line is set by
the local ion thermal speed $v_{th}$, the geometric width of the
local resonance with the radiation is about a Sobolev length,
$l_{Sob} \equiv v_{th}/(dv/dr)$ \citep{Sobolev}. Thus, in a
supersonic flow, this Sobolev length is of order $v_{th}/v \ll 1$
smaller than a typical flow variation scale, such as the
density/velocity scale height $H \equiv |\rho/(d\rho/dr)| \approx
v/(dv/dr)$. Therefore, while the flow speeds for the slow
acceleration solutions are lower than those of the steep
acceleration solutions, they are still supersonic throughout nearly
the entire wind, and so the Sobolev approximation remains quite
appropriate.

As noted in the introduction, it is important to keep in mind the
limited physical relevance of such 1D models. The 2D WCD simulations
of \citet{Owockietal94}, \citet{CO95}, and \citet{Owockietal96}
serve as an example, showing that for moderately rapid rotation,
there can be a 2D flow pattern by which material from higher
latitudes is focussed toward the equator through the WCD effect.
Depending on whether the material reaches the equator above or below
some ``stagnation point'', it either drifts outward or falls back
toward the star. This simultaneous infall plus outflow behavior is
not possible in a steady 1D model, but is a perfectly natural
occurrence in a 2D simulation. There are also the issues of
oblateness, limb darkening, and the non-radial line force, and how
these affect the latitudinal motion of the flow leading to
inhibition of the WCD \citep{CO95, Owockietal96, PP00}. Finally,
equatorial gravity darkening can reduce the wind mass flux from the
equator and lead to an equatorial wind density that is lower than
near the poles \citep{Owockietal96}. A key point here is that 1D
simulations represent a sort of ``best-case" scenario for the
formation of a disk. A move to 2D simulations shows that when the
relevant physics (such as nonzero, nonradial line forces and gravity
darkening) are included, material will tend to be channeled away
from the equator and inhibit disk formation. Thus, if 1D simulations
are incapable of producing equatorial densities capable of
explaining those inferred in B[e] supergiants, then 2D simulations
will most likely not change this.

Beyond even developing a 2D CAK-type model, there remain key
physical limitations not accounted for in this CAK formalism. Two
examples are the intrinsic, small-scale instability of line-driving,
and multiple scattering effects. The former might well disrupt a
slow-acceleration solution, even though the CAK form of such
solutions seem from the present simulations to be a stable
attractor. The latter places strict upper limits on the mass flux
that can be driven within a geometrically thin disk, since even with
ideally tuned line-driving parameters within a CAK model (e.g.
choosing an anomalously small CAK exponent $\alpha$ to enhance the
expected CAK-type mass loss rate), radiative driving in the thin
disk would generally be limited to one or two scatterings before the
photon escapes out of the disk plane \citep{Owocki06}. Thus, quite
generally, radiative driving could not produce a disk outflow that
exceeds the single scattering limit. So, to the extent that
observational inferences of supergiant B[e] disks imply a very dense
medium (dense enough to form dust), it seems more likely that these
represent orbiting \emph{Keplerian} disks, with perhaps ablation
flows off the disk surface producing the equatorial outflow inferred
by observations of Doppler-shifted absorption troughs in UV
resonance lines.

\acknowledgements{ We thank M. Cur{\'e} and R. Nikutta for
stimulating our initial interest in this topic. We thank J.
Blondin for continuing to make available the VH-1 hydrodynamics
code. This work was done with partial support of NSF grants
AST-0097983 and AST-0507581, NASA Chandra grant GO3-3024C, and DFG
project FE 573/3-1. }

\appendix

\section{Outer Critical Point Location for Shallow Solutions \\
with Finite Sound Speed}

Let us examine here a key issue regarding the location of the
outer critical point for the shallow acceleration solutions that
appear for rotation rates $\omega > 0.75$. For the zero
sound-speed limit (as analyzed in \S\S 2-3), the radius of this
critical point formally approaches infinity,
$r_{c} \rightarrow \infty$, or
$x_{c} \rightarrow 1$ in the scaled inverse radius coordinate
$x \equiv 1-\Rstar/r$.
This raises the issue of whether the numerical
hydrodynamic simulations (\S\S 4-5) done
with a finite outer boundary radius must use some sort of subcritical
right boundary condition (with one inward  and one outward pointing
characteristic), instead of the supercritical outflow
boundary conditions (with two outward characteristics) that are
usually used in stellar wind simulations.

Note however that the hydrodynamic simulations presented above
include a small but non-zero sound speed, characterized by the
dimensionless parameter $w_{s} \equiv a^{2}/v_{\rm esc}^{2} \approx
4 \times 10^{-4}$. When this is included in the equation of motion
(as in eq. [\ref{ScaledEoM}]), the additional term on the right side
has the scaled form $4 w_{s}/(1-x)$ (we are neglecting the
additional $w_{s}/w$ term on the left hand side of eq.
[\ref{ScaledEoM}]). The total effective gravity used in the
denominator of the nozzle function (see eq. [\ref{nozdef}]) then
takes the form \beq g(x) = 1 - \omega^{2} (1-x) - { 4 w_{s} \over
1-x} \, . \eeq
Instead of declining monotonically as $x$ increases, this now has a
maximum at some finite radius, and that in turn allows the nozzle
function to reach a minimum at some finite radius. In a model with a
non-zero sound speed, this thus becomes the effective critical point
of the slow acceleration solutions. Since $w_{s} \ll 1$, this
minimum still occurs near $x \ltwig 1$, and so to evaluate the
location, it is convenient to expand the finite disk correction
factor about $x=1$. For a $\beta=1$ velocity law and $\alpha=1/2$,
this gives \beq f(x) \approx 1 + \frac{1-x}{4} ~~ ; ~~ x \rightarrow
1 \, , \eeq which makes the nozzle function \beq n(x) \equiv {
f^{2}(x) \over g(x) } \approx { 1 + (1-x)/2 \over 1 - \omega^{2}
(1-x) - 4w_{s}/(1-x)} \, . \eeq We can easily solve for the critical
location $x_{c}$ through \beq n'(x_{c}) = 0 \,  . \eeq To lowest
order in the small parameter $\sqrt{w_{s}} \approx  0.02$, we find
\beq x_{c} \approx 1 - 2 \sqrt{ w_{s} \over \omega^{2} + 1/2}
\approx 1 -{ 0.04 \over \sqrt{\omega^{2} + 1/2}} \approx 1 - 0.04 \,
\sqrt{2} \, (1- \omega^{2}) \, , \eeq which implies a critical
radius of \beq {r_{c} \over \Rstar} \approx 17.67 \, (1 +
\omega^{2}) \, . \eeq

The upshot of this is thus that the outer boundary radius used in
our numerical models ($R_{\rm max} = 100 \Rstar$) should be well
above the critical point ($ r_{c} \ltwig 36 \Rstar$) for even the
shallow slope solutions at high rotation rate. Thus a pure
super-critical outflow boundary condition should indeed be
appropriate for all the models computed here.


\bibliographystyle{aastex}

\end{document}